\def\Granada{Instituto Carlos I de F\'{\i}sica Te\'{o}rica y Computacional,
Facultad de Ciencias, Universidad de Granada, Campus de Fuentenueva,
Granada 18002, Spain}

\def\INFN{Dipartimento di Scienze Fisiche and INFN, Sezione di Napoli,
      Mostra d'Oltremare, Pad. 19, 80125, Napoli, Italia.}

\def\Comision{Work partially supported by the Direcci\'{o}n General de 
              Ciencia y Tecnolog\'{\i}a.}
\def\IAA{Instituto de Astrof\'{\i}sica de Andalucia, Apartado Postal 3004,
         18080 Granada, Spain.}

\def\medio{\frac{1}{2}}
\def\imedio{\frac{i}{2}}
\def\nn{\nonumber}

\def\ni{\noindent}

\def\GC{{\cal G}_C}
\def\XL{\tilde{X}^L}
\def\XR{\tilde{X}^R}

\def\Gt{$\tilde{G}\,$}
\def\Gtm{\tilde{G}}
\def\z{\zeta}
\def\Xo{\tilde{X}_0}
\def\x{\vec{x}}
\def\q{\vec{q}}
\def\p{\vec{p}}
\def\a{\vec{a}}
\def\[{\left[}
\def\]{\right]}

\def\be{\begin{equation}}
\def\ee{\end{equation}}
\def\bea{\begin{eqnarray}}

\newcommand{\parcial}[1]{ \frac{\partial}{\partial #1} }

\documentstyle[12pt]{article}

\title{ HIGHER-ORDER QUANTIZATION ON A LIE GROUP
           \thanks{\Comision} }

\author{ V. Aldaya$^{1,2}$, J. Guerrero$^{1,2,3}$,
         and G. Marmo$^{3}$}

\date{November, 1998}

\begin{document}

\footnotetext[1]{\IAA} \footnotetext[2]{\Granada}
\footnotetext[3]{\INFN}

\maketitle

\begin{center} {\bf Abstract} \end{center}

In this paper we are mainly concerned with the study of
polarizations (in general of higher-order type) on a connected Lie
group with a $U(1)$-principal bundle structure. The representation technique
used here is formulated on the basis of a group quantization formalism
previously introduced which generalizes the Kostant-Kirillov co-adjoint
orbits method for connected Lie groups and the Borel-Weyl-Bott representation
algorithm for semisimple groups. We
illustrate the fundamentals of the group approach with the help of some
examples like the abelian group $R^k$ and the semisimple group $SU(2)$,
and the use of higher-order polarizations with the harmonic oscillator
group and the Schr\"{o}dinger group, the last one constituting the simplest
example of an anomalous group. Also, examples of infinite-dimensional anomalous
groups are briefly considered.

\vfil\eject

\section{Introduction}

This paper is devoted mainly to the concept of higher-order polarization
on a Lie group as a powerful tool in search of irreducibility of the
representations of the group, and/or the irreducibility of quantizations
in a group-theoretic quantization approach, in those {\it anomalous} cases
where standard (geometric) methods do not succeed.
The co-adjoint orbit method has proven to fail in quantizing, for instance,
non-K\"{a}hler orbits of certain Lie groups or orbits without any invariant
(first-order or standard) polarization. Also, the configuration-space image of
quantization of rather elementary physical systems cannot be obtained in a
natural way via the Geometric Quantization technique, but it is obtained
easily with the aid of higher-order polarizations.

Higher-order polarizations are naturally defined on the (left) enveloping
algebra of a Lie group as higher-order differential operators acting on
complex functions on the group manifold.
Since higher-order polarizations do not coincide, in general, with the
enveloping algebra of any first-order
polarization subalgebra, the space of wave functions is not necessarily
associated with any particular classical configuration space, thus leading
to {\it a breakdown of the notion of classical limit} for those systems
which are anomalous (see Sec. 4). It should be remarked that higher-order
polarizations excludes non-differential operators, such as
discrete transformations (like parity o charge conjugation), so that a
higher-order quantization on a group does not guarantee a full reduction of
the representation in the general case; it can lead, for instance, to the
direct sum of irreducible representations which are only distinguished by
the eigenvalues of a discrete operator commuting with the representation (see
the example of Sec. \ref{Algebraic Anomalies}). A generalization of the
concept of polarization including this sort of non-local operators is
beyond the scope of the present paper, but will be studied elsewhere.

The first general attempt to look at the problem of quantization in terms
of groups of transformations was done by L. van Hove \cite{Hove}, who tried
to associate a group of unitary transformations with the group of canonical
transformations on the phase space
$M=R^{2n}$. The main point in van Hove's approach was to
consider a Hilbert space of complex valued functions on the phase space $M$,
and to implement on these wave functions the action of canonical
transformations as unitary transformations according to
\begin{eqnarray}
&1)& \phi\rightarrow U_{\phi} \label{phase} \\
&2)& (U_{\phi}f)(p,q)=(e^{-iS_{\phi}}f)(P,Q) \nn\,,
\end{eqnarray}

\ni where $S_{\phi}$ is the
generating function of the canonical transformation
$\phi:(P,Q)\rightarrow (p,q)$:
\begin{equation}
p_idq^i-P_idQ^i = dS_{\phi}\,.
\end{equation}

\ni When the transformation $\phi$ is a one-parameter group
of transformations an infinitesimal relation
for the generating function can be associated:
\begin{equation}
L_X(p_idq^i)=dS_X \,,
\end{equation}

\ni where $X$ is the infinitesimal generator of the canonical
one-parameter group $\phi_t$. It should be stressed that because
of the phase factor in front of $f$ in (\ref{phase}), this action does not
respect the ring structure (it is not an automorphism) of ${\cal F}(M)$,
indeed $U_\phi(f_1{\cdot}f_2)\neq U_\phi(f_1){\cdot}U_\phi(f_2)$.

Furthermore, in the known duality relation between a manifold
$M$ and the ring of (germs of) functions on it, ${\cal F}(M)$,
a one-to-one correspondence between automorphisms of ${\cal F}(M)$
and diffeomorphisms on $M$ can be established. Therefore, if $U_{\phi}$
were an automorphism of the ring of functions when $\phi$ is a
one-parameter group, a vector field could be associated with it, and this
would coincide with the infinitesimal generator of $U_{\phi}$ as from the
Stone-von Neumann theorem \cite{Stone}, after a measure would have been
introduced associated with the canonical symplectic structure.

It should be realized that $U_{\phi}$ is a ring automorphism (up to a
redefinition to absorb a global phase) only if
$dS_{\phi}=0$ or, in equivalent terms, $dS_X=0$. To achieve
this result we may enlarge the carrier space
$M$ to $\tilde{M}=M\times U(1)$. We use 
$\pi$ to denote
the projection
$\pi:\tilde{M}\rightarrow M$, and define $\tilde{\theta}=\pi^*(p_idq^i)
-ds$, $s$ being the extra coordinate, and $\tilde{X}=X+
S_X\frac{\partial}{\partial s}$. Then
\begin{equation}
L_{\tilde{X}}\tilde{\theta}= \pi^*(L_X(p_idq^i))-dS_X=0\,.
\end{equation}

\ni Thus, the lift of our one-parameter group of transformations $U_{\phi}$
to the extended space would be a ring automorphism and, therefore, associated
with an infinitesimal generator which is a vector field on $\tilde{M}$.
This is a possible motivation for introducing a $U(1)$ ``extension'' of our
symplectic manifold $M$. Geometric Quantization \cite{Souriau,Woodhouse} 
generalizes this construction
to any symplectic manifold $M$ with symplectic 2-form of integer cohomology
class. In the general case, the principal fibration 
$\tilde{M}\rightarrow M$ with structure group $U(1)$ is not necessarily 
trivial.

After this motivation, in Sec. 2 the group approach to quantization is
introduced, illustrated with the examples of the quantization of the abelian
group $R^k$, leading to the Heisenberg-Weyl commutation relations, and the
quantization of the semisimple group $SU(2)$, in Sec. 3.
The notion of (algebraic) anomaly is discussed  and the particular cases
of the Schr\"{o}dinger and Virasoro groups are considered in Sec. 4.
In Sec. 5 we provide a precise definition of higher-order polarizations
as well as examples of the use of higher-order polarizations for the
case of the quantization of the (non-anomalous) harmonic oscillator group in
configuration space and the Schr\"{o}dinger group. Finally, Sec. 6 is devoted
to comments and outlooks.
The definitions of both first-order and higher-order polarizations given in
this paper generalize in some respects the ones introduced in previous
papers (see \cite{23,Anomalias}).

\section{Group Approach to Quantization (GAQ)}

Now we are no longer dealing necessarily with a symplectic manifold,
but rather with a Lie group $\tilde{M}=\tilde{G}$. The discussion in the
Introduction represents the motivation for starting with a Lie group  which is
a $U(1)$-bundle with
base manifold $M$, in particular, with a central extension $\tilde{G}$ of
$G=M$. From the physical point of view, this is traced back to the fact that
the Hilbert space of physical states is made out not of functions, but of rays,
that is, equivalence classes of functions differing by a phase. This allows
for representations of $G$ which are projective representations, and these
are true representations of certain central extensions \Gt of G.

We shall restrict ourselves to the study of groups $G$ which are connected
and simply connected. This is by no means a drawback of our formalism, since 
for a non-simply connected group $G$, if $\bar{G}$ is
the universal covering group and $p:\bar{G}\rightarrow G$ is the canonical
projection, a projective unitary irreducible representation $U$ of $\bar{G}$
is also a projective unitary irreducible representation of $G$ if and only
if $U(\bar{g})=e^{i\alpha}I\,,\forall \bar{g}\in {\rm ker}p$. In this way, 
we can obtain easily the representations of $G$ if we know the representations
of its universal covering group $\bar{G}$.

The connection 1-form $\tilde{\theta}\equiv \Theta$ on $\tilde{G}$ will be
naturally selected (see below) among the components of the left-invariant,
Lie Algebra valued, Maurer-Cartan 1-form, in such a way that
$\Theta(\Xo)=1$ and $L_{\Xo}\Theta=0$, where $\Xo$ is the vertical
(or fundamental) vector field. $\Theta$ will play the role of a
connection 1-form associated with the $U(1)$-bundle structure.

Therefore, let us suppose that $G$ is a connected and simply connected Lie
group and \Gt is a central extension of $G$ by $U(1)$. On \Gt we shall
consider the space of complex functions $\Psi$ satisfying
the $U(1)$-equivariance condition, i.e. $\Psi(\zeta*g)=\zeta\Psi(g)$, on
which the left translations of $\Gtm$ (generated by right-invariant vector
fields $\tilde{X}^R$) realize a representation of the group which is in general
reducible. Fortunately, right translations (generated by left-invariant
vector fields $\tilde{X}^L$) on any Lie group do commute with left ones, so 
that polarization subgroups $G_{\cal P}$ can be sought (see below) to perform 
a (maximal) reduction of the representation through polarization conditions,
defining a subspace of wavefuntions ${\cal H}$ satisfying 
$\Psi(g*G_{\cal P})=\Psi(g)$ (or $L_{\tilde{X}^L}\Psi=0,\ \forall
\tilde{X}^L\in {\cal P}$,
where ${\cal P}$ is the Lie algebra of $G_{\cal P}$).
For a rather wide class of groups, polarization subalgebras
in the Lie algebra suffice to achieve the full reduction, whereas groups
called anomalous (precisely those for which the Kirillov-Kostant methods is
not appropriate, see Sec. 4) require generalized or higher-order polarization
subalgebras of the enveloping algebra of $\Gtm$, and even so, some special
representations can be only reduced to a direct sum of irreducible
representations distinguished by the eigenvalues of a discrete operator
not belonging to the group (see Sec. 5).

Finally, the Hilbert space structure on the space of wave functions is
provided by the invariant Haar measure constituted by the exterior product of
the components of the left-invariant canonical 1-form, 
$\Omega^L\equiv\theta^{Lg^1}\wedge\theta^{Lg^2}\wedge...$. On the reduced 
space (unless for the case of first order polarizations), the existence of a 
quasi-invariant measure $\mu$ is granted, since the wave functions have 
support on $G/G_{\cal P}$, 
which is a homogeneous space (see \cite{Barut}, for instance). This means that
the operators $X^R$, once reduced to ${\cal H}$, are (anti-)hermitian with 
respect to the scalar product defined by $\mu$, or they can be made 
(anti-)hermitian by the addition of the Radon-Nikodym derivative in the case
where $\mu$ is only quasi-invariant. A detailed study of the invariance
properties of measures on the space of polarized wave functions is in
preparation (see \cite{Unitariedad}).


The classical theory for the system is easily recovered by defining the
Noether invariants as $F_{g^i}\equiv i_{\XR_{g^i}} \Theta$. A Poisson
bracket can be introduced (defined by $d\Theta$), in such a way that the
Noether invariants generate a Lie algebra isomorphic (if there are no
algebraic anomalies, see Sec. 4) to that of \Gt (see \cite{23} for a complete
description of the classical theory using the formalism of GAQ).

\subsection{$U(1)$-central extensions and connection 1-form}

To consider the quantization procedure \cite{23,Ramirez}, in particular to
address the definition of the quantization 1-form $\Theta$ and to exhibit
the vector fields which are the generators of the desired
$\tilde{U}_{\phi}$, we shall be more specific and state our hypothesis more
precisely.

Following Bargmann \cite{Bargmann}, instead of characterizing central 
extensions by smooth 2-cocycles ({\it factors}) 
$\omega: G\times G \rightarrow U(1)$, we shall employ 
{\it exponents} $\omega=e^{i\xi}$, i.e. 2-cocycles on $G$ with 
values in $R$, $\xi:G\times G\rightarrow R$. In addition, for a simply
connected group, any local factor or exponent defined on a neighborhood of 
the identity
can be extended to a global factor or exponent defined on the whole group
\cite{Bargmann}.

In order to select, from the very beginning, the particular projective
unitary irreducible representation of $G$ that we want to obtain, we shall
consider the central extension \Gt characterized not only by an
equivalence class of 2-cocycles $[[\xi]]$ on $G$ (that is, the set of
2-cocycles $\xi:G\times G\rightarrow R$ which differ among them by a
coboundary, i.e. a 2-cocycle $\xi_\lambda(g_1,g_2)$ generated by a function
$\lambda:G\rightarrow R$, $\xi_\lambda(g_1,g_2)=
 \lambda(g_1*g_2)-\lambda(g_1)-\lambda(g_2)$) \cite{Bargmann}, but by a
particular
subclass $[\xi]$ of the class $[[\xi]]$. We proceed in this fashion
because, as we shall see later, each cohomology class $[[\xi]]$ can be
subsequently partitioned into {\it pseudo-cohomology} classes, which are
essentially associated with (classes of) coboundaries generated by functions
$\lambda:G\rightarrow R$ with non-zero gradient at the identity $e$ of $G$.

This implies, in particular, that we can select for $\xi(g_1,g_2)$ a pure
coboundary (a 2-cocycle cohomologous to the trivial 2-cocycle
$\xi_0(g_1,g_2)\equiv 0$), but belonging to a non-trivial pseudo-cohomology
class of coboundaries generated by a function $\lambda$ with non-zero gradient
at the identity. This will be the case, for instance, for (the universal 
covering groups of) all semisimple
(finite-dimensional) groups \cite{Jacobson} and the Poincar\'{e} group
(see Refs. \cite{Souriau}
and \cite{Position}), all of them with trivial $2^{nd}$ cohomology
group. The advantage of this procedure is that it will allow us to obtain all 
the unirreps that are included in the regular representation by using 
the same technique. We should remark that $H^2(G,U(1))$ in the Bargmann
cohomology \cite{Bargmann} corresponds to the first standard-cohomology group
$H^1(G,{\cal G}^*)$ (see, for instance, Refs. \cite{Mickelsson,Souriau}) of $G$
with values on the co-adjoint module ${\cal G}^*$, which is in turn equivalent
to the first Cartan-Eilenberg cohomology group \cite{Jacobson}.

 Therefore, we shall consider a quantization group \Gt as a central extension
of $G$, connected and simply connected, by $U(1)$ characterized by a 2-cocycle
(or exponent) $\xi:G\times G\rightarrow R$, satisfying: 
\begin{eqnarray}
\xi(g_1,g_2) + \xi(g_1*g_2,g_3) &=& \xi(g_1,g_2*g_3) + \xi(g_2,g_3) \nn\\
\xi(e,e)&=&0 \,.  \label{cociclo}
\end{eqnarray}

\ni Most of the 
following construction also applies to the case
in which $\Gtm$ is a non-trivial principal bundle on $G$, so that a local
2-cocycle must be defined on each fibre-bundle chart. This is the case of
Kac-Moody groups (see for instance \cite{Pressley}).


The group law for \Gt\  can be written (since in all the cases we are
considering \Gt\  is the trivial topological product of $G$ and $U(1)$, the
2-cocycle $\xi$ will be smooth) as:
\be
(g',\z')*(g,\z)=(g'*g,\z'\z e^{i\xi(g',g)})\, .
\ee

Considering a set of local coordinates at the identity
 $\{g^i,\,\, i=1,\ldots, {\rm dim} G\}$ in $G$, the group
law can be written as the
set of functions  
$g''{}^i=g''{}^i(g'{}^j,g^k)$, $j,k=1,\ldots,{\rm dim} G$.
We introduce the sets of left- and right-invariant vector fields of \Gt\ 
associated with the set of coordinates $\{g^i\}$ as those which
 are written as $\parcial{g^i},\,\,i=1,\ldots,{\rm dim} G$ at the identity,
that is:
\be \XL_{g^i}(\tilde{g}) = \tilde{L}^T_{\tilde{g}} \parcial{g^i}\,,
\qquad \XR_{g^i}(\tilde{g}) = \tilde{R}^T_{\tilde{g}} \parcial{g^i}\,,
\ee

\ni where the tilde refers to operations and elements in \Gt. The left- (and
right- since the U(1) subgroup is central in \Gt) invariant vector field
which at the identity is written as $\parcial{\phi}$, with $\phi=-i\log\z$, is
$\tilde{X}_{\z}(\tilde{g})\equiv 2{\rm Re}(i\z\parcial{\z})=\parcial{\phi}$. 
We usually keep the sub/superscript
$\z$ instead of $\phi$ to remind the reader that we are dealing with the 
compact
fibre $U(1)$ and not $R$. $\Xo\equiv\tilde{X}_{\z}$ is the vertical (or
fundamental) vector field associated with the fibre bundle
\be   U(1) \rightarrow \Gtm \rightarrow G \,.
\ee

Analogous considerations can be made for the sets of left- and right-invariant
1-forms associated with the set of local coordinates $\{g^i\}$,
i.e. those which at the identity are written as $dg^i$:
\be
\tilde{\theta}^L{}^{g^i}(\tilde{g})=\tilde{L}^*_{\tilde{g}^{-1}}dg^i =
 \theta^L{}^{g^i}(g) \,,
\qquad
\tilde{\theta}^R{}^{g^i}(\tilde{g})=\tilde{R}^*_{\tilde{g}^{-1}}dg^i =
\theta^R{}^{g^i}(g)\,.
\ee

For simplicity of notation, we shall omit the point in which the vector
fields and 1-forms are calculated. Due to the left and right invariance, we
have $\theta^L{}^{g^i} (\XL_{g^j})=\delta^i_j = \theta^R{}^{g^i}(\XR_{g^j})$.

We can also compute the left- and right-invariant 1-forms which are dual to
the vertical generator $\Xo$:
\be
\tilde{\theta}^{L\z}=\tilde{L}^*_{\tilde{g}^{-1}} d\phi \,,
\qquad
\tilde{\theta}^{R\z}=\tilde{R}^*_{\tilde{g}^{-1}} d\phi \,.
\ee

%

We shall call $\Theta\equiv \tilde{\theta}^L{}^{\z}=\frac{d\z}{i\z} +
 \frac{\partial \xi(g',g)}{\partial g^i}|_{g'=g^{-1}} dg^i$,
 where $\frac{d\z}{i\z}=d\phi$, the
{\it quantization 1-form}. It defines a connection on the fibre bundle \Gt
and it is uniquely determined by the 2-cocycle $\xi(g_1,g_2)$ (it does not
change under changes of local coordinates of $G$). 

Adding to $\xi$ a coboundary $\xi_{\lambda}$, generated by the function
$\lambda$, results in a new group (note that the new group
$\Gtm'$ is isomorphic to \Gt, the isomorphism being 
$(g,\z)\mapsto (g,\z e^{i\lambda})$) 
 $\Gtm'$ and a new quantization 1-form 
$\Theta' = \Theta +\Theta_{\lambda}$ with
\be
\Theta_{\lambda} = \lambda^0_i \theta^L{}^{g^i} - d\lambda \,,
\ee

\ni where $\lambda^0_i\equiv \frac{\partial \lambda(g)}{\partial g^i}|_{g=e}$,
that is, the gradient of $\lambda$ at the identity with respect to a set of
local canonical coordinates \cite{Bargmann} (see also \cite{Pontrjaguin}).
Note that $\Theta$ is left-invariant under \Gt and $\Theta'$ is left-invariant
under $\Gtm'$, therefore $\Theta_{\lambda}$ is neither invariant under \Gt nor
$\Gtm'$. However, since
$\lambda^0_i$ are constants, up to the total differential $d\lambda$, 
$\Theta_{\lambda}$ is
left invariant under $G$ (and consequently under \Gt and $\Gtm'$). 
We also have $d\Theta_\lambda = \lambda^0_i
d\theta^L{}^{g^i}$, so that using the relation
\be
d\theta^L{}^{g^i} = \medio C_{jk}^i \theta^L{}^{g^j} \wedge \theta^L{}^{g^k}\,,
\ee

\ni where $C_{jk}^i,\,i,j,k=1,\ldots, {\rm dim} G$ are the structure
constants of the Lie algebra ${\cal G}\equiv T_e G$ in the basis of the
left-invariant
vector fields associated with the set of local canonical coordinates
$\{g^i\}$, we obtain:
\be
d\Theta_{\lambda} =
\medio \lambda^0_i C_{jk}^i \theta^L{}^{g^j}\wedge \theta^L{}^{g^k}\,.
\ee

Note that $\lambda$ defines an element $\vec{\lambda}^0$ of the coalgebra
${\cal G}^*$ of $G$ characterizing the presymplectic form
$d\Theta_{\lambda}\equiv d\Theta_{\lambda^0}$. It is easy to see then that
given $\vec{\lambda}'{}^0$ and $\vec{\lambda}^0$ on the same orbit of the
coadjoint action of $G$, $\vec{\lambda}'{}^0= Ad(g)^*\vec{\lambda}^0$, for
some $g\in G$, the corresponding presymplectic forms are related through:
\be
d\Theta_{\lambda'{}^0} \equiv d\Theta_{Ad^*(g)\lambda^0} =
Ad^*(g) d\Theta_{\lambda^0}
\ee

 Taking into account that $(G/G_{\lambda^0},d\Theta_{\lambda^0})$, where 
$G_{\lambda^0}$ is the isotropy group in the point $\vec{\lambda}^0$ under the 
coadjoint action of $G$, is a 
symplectic manifold symplectomorphic to the coadjoint orbit through 
$\vec{\lambda}^0$ with its natural symplectic 2-form  
$\omega_{\lambda^0}$, and using proposition 4.4.2 of \cite{Pressley}, 
we can say that $\xi_\lambda$ is a well-defined 2-cocycle if and only if
$\omega_{\lambda^0}$ is of integral class \cite{Kirillov}.

In summary, we can classify the central extensions of $G$ in equivalence
classes, using two kinds of equivalence relations, one subordinated to the
other. The first one is the
standard one which leads to the $2^{nd}$ cohomology group $H^2(G,U(1))$,
where two 2-cocycles are cohomologous if they differ by a coboundary.
According to this we associate with $\xi$ the class $[[\xi]]$, the elements
of which differ in a coboundary generated by an arbitrary function on $G$.
With this equivalence class we can associate a series of parameters, given
by the corresponding element of $H^2(G,U(1))$, and will be called the {\it
cohomology parameters}. An example of them is the mass parameter which
characterizes the central extensions of the Galilei group. However, the
previous considerations suggest that each equivalence class $[[\xi]]$
should be further partitioned according to what we shall call {\it
pseudo-cohomology classes}, $[\xi]$, the elements of which differ by a
coboundary $\xi_{\alpha}$ generated by a function $\alpha$ on $G$ having
trivial gradient at the identity. Pseudo-cohomology classes are then
characterized by coadjoints orbits of ${\cal G}^*$ which satisfy the
integrallity condition (the condition of integrallity is
associated with the globallity of the generating function $\lambda$ on the
group).  With these pseudo-cohomology classes we associate a series of 
parameters called the {\it pseudo-cohomology parameters}. An example of 
this is the spin for the Galilei group or the spin and the mass for the 
Poincar\'e group. Note that these parameters are associated with (integral) 
coadjoint orbits of the correspondig groups.

The idea of subclasses inside $H^2(G,U(1))$ was firstly introduced
by Saletan \cite{Saletan} who noted that under an In\"{o}n\"{u}-Wigner contraction
some coboundaries $\xi_{\lambda}$ of $G$ become non-trivial cocycles
$\xi_c$ of the contracted group $G_c$ since the generating function
$\lambda$ is badly behaved in the contraction limit while $\xi_{\lambda}$
itself has a well defined limit $\xi_c$. The simplest physical example is
that of the Poincar\'{e} group whose pseudo-cohomology group goes to the
cohomology group of the Galilei group \cite{Pseudo}. For semisimple groups,
pseudo-cohomology is also related to the $\check{C}$ech  cohomology of the
generalized Hopf fibration by the Cartan subgroup $H$, $G\rightarrow G/H$,
\cite{Formal}.

 Some comments on pseudo-cohomology classes are in order. Firstly, in each
pseudo-cohomology class one can always find  representatives that are linear
in the coordinates $\{g^i\}$ in a neighbourhood of the identity, that is,
if $\xi_{\lambda}$ is a pseudo-cocycle with generating function $\lambda(g)$
having gradient  $\vec{\lambda}^0$ at the identity, then $\xi_{\lambda}$ is
pseudo-cohomologous to the pseudo-cocycle generated by 
$\lambda_i^0g^i$, since $\lambda(g)-\lambda_i^0g^i$ has zero gradient at the 
identity of the group.

Secondly, pseudo-cocycles $\xi_{\lambda}$ associated with those points 
$\vec{\lambda}^0\in {\cal G}^*$ that are invariant under the coadjoint action
of $G$ (i.e. they constitute zero-dimensional coadjoint orbits) are either 
zero or belong to the trivial pseudo-cohomology class, since for them 
$d\Theta_{\lambda}=0$. Therefore, zero-dimensional coadjoint orbits do not 
lead to central (pseudo-)extensions. This is reasonable since zero-dimensional 
coadjoints orbits are associated with one-dimensional representations, which 
are abelian. However, pseudo-cocycles associated with zero-dimensional 
coadjoint orbits, in certain cases, play an important role in the process of 
unitarizing representations (see \cite{Unitariedad}) for which an invariant 
measure on the representation space does not exist (and only quasi-invariant 
ones can be found). The unitarization process involves central 
pseudo-extensions by the abelian multiplicative group $R^+$ leading to the
Radon-Nikodym derivative.

Finally, the quantum operators $\XR_i$ obtained from the quantization of a
pseudo-extended group should be redefined with the addition of the linear
terms $\lambda^0_i\Xo$, i.e. $\XR_i{}' \equiv \XR_i+\lambda^0_i\Xo$, in 
order to obtain the original commutations relations (see the example of 
$SU(2)$ in Sec. 3.2).

In the general case, including infinite-dimensional semi-simple Lie groups,
for which the Whitehead lemma does not apply, the group law for
$\tilde{G}$ will contain cocycles as well as pseudo-cocycles
(see \cite{Virasoro,Formal,Anomalias,Mickelsson2}). The simplest
physical example of a quantum symmetry including such an extension is that
of the free non-relativistic particle with spin; the Galilei group must
be extended by a true cocycle to describe the canonical commutation relations
between
q's and p's as well as by a pseudo-cocycle associated with the
Cartan subgroup of $SU(2)$, to account for the spin (we shall consider
the universal covering of the Galilei group, so that we obtain SU(2) as the
rotation group and therefore half-integer values are allowed for the spin)
degree of freedom
\cite{Position}.

Let us therefore consider a Lie group \Gt which is a $U(1)$-principal
bundle with the bundle projection $\pi:\Gtm \rightarrow G$ being a group
homomorphism. We denote by $\Theta$ the connection 1-form constructed as
explained earlier. It satisfies $i_{\Xo}\Theta=1,\,\,L_{\Xo}\Theta=0$, with
$\Xo$ being the infinitesimal generator of $U(1)$, or the fundamental
vector field of the principal bundle, which is in the centre of the Lie
algebra $\tilde{\cal G}\equiv T_e \Gtm$. Since $\Theta$ is left-invariant
it will be preserved ($L_{\XR}\Theta=0$) by all right-invariant vector
fields (generating finite left translations) on $\tilde{G}$. These vector
fields are candidates to be infinitesimal generators of unitary
transformations. To define the space of functions on which they should act,
we proceed as follows. Choose a representation of the structure group
$U(1)$, which will be the natural representation on the complex
numbers, and build the space of complex functions on $\tilde{G}$ that satisfy
the $U(1)$-equivariance condition
\begin{equation}
L_{\Xo}\psi=i\psi\,. \label{equivariance}
\end{equation}

\ni  where $\Xo$ is the fundamental vector field on the principal
bundle $\tilde{G}\rightarrow G$.
This space is isomorphic to the linear space of sections of the
bundle $E\rightarrow \tilde{G}/U(1)$, with fibre $F\equiv C$, associated with
$\tilde{G}\rightarrow G$ through the natural representation of $U(1)$ on the
complex numbers $C$ (see, for instance, \cite{Bal}). 
To get an irreducible action of the
right-invariant vector fields  we have to select appropriate subspaces,
and this will be achieved by polarization conditions.

\subsection{Additional structures associated with the connection 1-form}

The 2-form $\tilde{\Sigma}\equiv d\Theta$ is left invariant under \Gt, and
is projectable to a left invariant 2-form $\Sigma$ of $G$.
This, evaluated at the identity,  defines a 2-cocycle on the Lie algebra
$\cal G$.


On vector fields ${\cal X}(G)$ we can define a
``generalized Lagrange bracket'' by setting, for any pair of vector fields
$X,Y\in {\cal X}(G)$,
\begin{equation}
(X,Y)_{\Sigma} = \Sigma(X,Y)\in {\cal F}(G)\,.
\end{equation}

In particular, when we consider left invariants vector fields
$X^L,Y^L\in {\cal X}^L(G)$, we get a  real valued bracket:
\begin{equation}
(X^L,Y^L)_{\Sigma} = \Sigma(X^L,Y^L)\in R\,.
\end{equation}

By evaluating $\tilde{\Sigma}$ at the identity of the group, i.e. on
$T_e \Gtm=\tilde{\cal G}$,
we can bring it to normal form which would be the analog of a Darboux frame in
the space of left-invariant 1-forms. We can write
\begin{equation}
\tilde{\Sigma} = \sum_{a=1}^{k}\theta^L{}^a\wedge\theta^L{}^{a+k}\,,
\end{equation}

\ni where $\theta^L{}^a,\theta^L{}^{a+k},\, a=1,\ldots,k$ are left-invariant 1-forms.
We can define a (1,1)-tensor field $J$, a partial (almost) complex structure,
by setting:
\begin{eqnarray}
J\theta^L{}^a &=& \theta^L{}^{a+k} \nn \\
J\theta^L{}^{a+k} &=& -\theta^L{}^a \\
J\theta^L{}^l &=& 0 \nn\, ,
\end{eqnarray}

\ni where $\theta^L{}^l$ are the remaining elements of a basis of left-invariant
1-forms not appearing in $\tilde{\Sigma}$ (that is, dual to vector fields in
$Ker\tilde{\Sigma}$.). We also have a ``partial metric tensor'' $\rho$ by
setting $\rho(\theta^L{}^a,\theta^L{}^{a'})=\delta_{aa'},\,
\rho(\theta^L{}^a,\theta^L{}^l)=0,\,\rho(\theta^L{}^l,\theta^L{}^{l'})=0$.

Our considerations will be always restricted to finite-dimensional Lie
groups or infinite-dimensional ones possessing a countable basis of generators
for which, for arbitrary fixed $\tilde{X}^L, \, \tilde{\Sigma}(\tilde{X}^L,
\tilde{Y}^L)=0$ except for a finite number of vector fields $\tilde{Y}^L$
(finitely non-zero cocycle), and therefore this partial (almost) complex
structure $J$ can always be introduced.

It is possible to associate with $\Theta$ an horizontal projector, a
(1,1)-tensor field. We first define the vertical projector
$V_{\Theta}(X)=\Theta(X) X_0$, and then $H_{\Theta}=I-V_{\Theta}$.

The characteristic module of $\Theta$ is defined to be the intersection
of $Ker\Theta$ and $Ker d\Theta=Ker\tilde{\Sigma}$. By restricting to
${\cal X}^L(\Gtm)$ we get the characteristic subalgebra $\GC$. Elements in
$Ker\Theta\cap Ker d\Theta$ are easily shown to be a Lie algebra. In fact,
it follows from the identity
$d\Theta(X,Y)=L_X\Theta(Y)-L_Y\Theta(X) -\Theta([X,Y])$.

It turns out that the quotient of $\Gtm$ by the integrable distribution
generated by ${\cal G}_C$, $P\equiv \Gtm/{\cal G}_C$, is a quantum
bundle in the sense of Geometric Quantization, with connection the projection
of $\Theta$ to $P$ (see \cite{23}). Therefore, $d\Theta$ projected onto
$P/U(1)$ is a symplectic 2-form, establishing the connection with the
Coadjoint Orbits Method, the different coadjoint orbits being obtained by
suitable choice of the (pseudo)extension parameters. To be more specific,
if $\Omega_{\mu}$ is the orbit through an arbitrary point 
$\mu\in {\cal G}^*$, we can construct the
left-invariant 1-form $\theta_{\mu}^L\in {\cal X}^L(G)^*$ and $\omega_{\mu}\equiv d\theta_{\mu}^L$.
Then, the quotient of $G$ by the characteristic distribution of $\omega_{\mu}$,
$G/Ker\{\omega_{\mu}\}$ is differentiably symplectomorphic to $\Omega_{\mu}$:
\begin{equation}
(\Omega_{\mu}\,,\,\mu.[\,,\,])\;\approx\;(G/Ker\{\omega_{\mu}\}\,,\omega_{\mu})\,,
\end{equation}

\ni where $\mu.[\,,\,]$ is the standard symplectic form on the coadjoint orbits.
Note that had we taken $\Gtm$ instead of $G$, $\mu.[\,,\,]$ would have been
replace by $\tilde{\mu}.[\,,\,]=\mu.[\,,\,]+\tilde{\Sigma}$. By considering
functions on this orbit, we have the possibility of considering them as functions on $G$
(by taking the corresponding pull-back). Now, to these functions we can
apply the left and right action of $G$ or, even, the operators in the enveloping
(left and right) algebra, and these operators are not available in ${\cal G}^*$.

However, we are not going to consider such a quotient explicitly. Rather,
the inclusion of the characteristic subgroup in the pre-contact manifold
$\Gtm$ represents a non-trivial improvement and generalization of Geometric
Quantization in the sense that equations of motion
can be naturally included into the quantization scheme, and also because
we are not forced, this way, to consider the classical equations of motion,
the solutions of which might be lacking.

\subsection{Polarizations}

As commented at the beginning of Sec. 2, to reduce the representation
constituted by the left action of the group \Gt on wave functions
satisfying (\ref{equivariance}), we need to select appropriate invariant
subspaces. This is achieved by means of the polarization conditions, in
terms of suitable subalgebras of left-invariant vector fields. Thus,
a {\bf first-order polarization} or just {\bf polarization} ${\cal P}$ is
defined as a maximal horizontal left subalgebra. The horizontality
condition means that the polarization is in $Ker\Theta$. Again, by using
the identity $d\Theta(X,Y)=L_X\Theta(Y)-L_Y\Theta(X) -\Theta([X,Y])$ we
find that the generalized Lagrange bracket  of any two elements of ${\cal
P}$ vanishes. Therefore we find that a polarization is an isotropic maximal
subalgebra. We notice that maximality is with respect to the Lie commutator
(subalgebra) not with respect to isotropy (Lagrange bracket).

A polarization may have non-trivial intersection with the characteristic
subalgebra.
We say that a polarization is {\bf full (or regular)} if it contains the
whole characteristic subalgebra.
We also say that a polarization $\cal P$ is
{\bf symplectic} if $\tilde{\Sigma}$ on ${\cal P} \oplus J{\cal P}$ is
of maximal rank. Full and symplectic polarizations correspond to
{\it admissible} subalgebras subordinated to $\Theta|_e\in \tilde{\cal G}^*$
\cite{Kirillov}.


It should be stressed that the notion of polarization and characteristic
subalgebras here given in terms of $\Theta$ is really a consequence
of the fibre bundle structure of the group law of \Gt and, therefore,
can be translated into finite (versus infinitesimal) form defining
the corresponding subgroups (see \cite{Ramirez}).

From the geometric point of view, a polarization defines
a foliation via the Frobenius theorem. It is possible to select subspaces of
equivariant complex valued functions on \Gt, by requiring them to be constant
along integral leaves of the foliation associated with the polarization.
Whether this subspace is going to carry an irreducible representation
for the right-invariant vector fields is to be checked. When the polarization
is full and symplectic we get leaves which are maximally isotropic
submanifolds for $d\Theta$. The selected subspaces of equivariant complex
valued functions on \Gt, which we may call wave functions, will be
characterized by $L_{\tilde{X}_0}\Psi=i\Psi,\, L_{X^L}\Psi=0,\,\forall X^L\in
{\cal P}$.

{\it Remark:} We can generalize the notion of polarization by simply relaxing the
condition of horizontality, and defining a {\bf non-horizontal polarization}
as a maximal left subalgebra not containing the vertical generator $\Xo$.
Although this kind of polarizations are not horizontal with respect to the
quantization 1-form $\Theta$, it is always possible to find a new $\Theta'$
for which a given non-horizontal polarization becomes horizontal, and
$\Theta'$ is of the form:
\be
\Theta' = \Theta + \alpha_i \theta^L{}^{g^i} \,,
\ee

\ni which implies that, up to a total differential, $\Theta'$ is obtained
adding a coboundary (pseudococycle) to the original 2-cocycle. Therefore,
the description in terms of pseudo-extensions and that of non-horizontal
polarizations are equivalent.

\section{Simple examples}
To see how this construction works, let us consider in detail two paradigmatic examples:
the abelian group $R^k$ and the semisimple one $SU(2)$. These suffice to illustrate
GAQ in its easiest (first-order) form. Anomalous cases will be encountered in
the next sections.


\subsection{The abelian group $R^k$}


Despite of the fact that $R^k$ is the simplest locally compact group, it
deserves a detailed study since its non-trivial (projective) representations,
the group being non-compact, are infinite-dimensional, leading to the
well-known Heisenberg-Weyl commutation relations, the base for non-relativistic Quantum Mechanics.

 We shall parameterize $R^k$ by (global in this case) canonical coordinates
$\x=(x^1,\ldots,x^k)$. Since the group is abelian, the coadjoint action is
trivial and its coadjoint orbits are points (zero-dimensional). Therefore,
there will not be pseudo-cohomology classes and only the cohomology group
is relevant. This means that given any 2-cocycle $\xi$ defining a central
extension $\tilde{R}^k$ of $R^k$, $\Theta-\frac{d\z}{i\z}$, where $\Theta$ is
the quantization 1-form, is not left invariant under the action of $R^k$, that
is, under translations (not even up to a total differential). Since
$d\Theta$ is always left invariant, we see that it is an exact 2-form but
it is not invariantly exact, this fact being a consequence of the
non-trivial group cohomology of $R^k$. Since the group is abelian,
$d\Theta$ takes the same value at all points, and can be written as:
\be  d\Theta= a_{ij} dx^i\wedge dx^j\,,
\ee

\ni where $a_{ij}$ is an antisymmetric $k\times k$ matrix. This allows us
to write any 2-cocycle in the form:
\be \xi(\x_1,\x_2) = a_{ij} x^i_1 x^j_2\,,
\ee

\ni up to a coboundary which, due to the trivial pseudo-cohomology of $R^k$,
will always contribute to $\Theta$ with an irrelevant total differential.
Therefore, $\xi$ is an antisymmetric bilinear function on $R^k$, and with
an appropriate change of coordinates in $R^k$ can be taken to normal form,
in which the matrix $a_{ij}$ is written as:
\be
\medio \left( \begin{array}{ccccc}
0_n& |& D_n &|& \\
---&  & --- & & \vec{0}_{2n} \\
-D_n &|& 0_n &|& \\
---& & --- & & --- \\
 & ^t\vec{0}_{2n} & & |& 0_r
\end{array} \right)\,,
\ee

\ni where $0_p$ is the $p\times p$ zero matrix, $\vec{0}_{2n}$ is the zero
$2n$-dimensional column vector, and $D_n$ is a $n\times n$ real matrix of
the form:
\be \left( \begin{array}{cccccc}
 \nu_1 & 0 & . & .& .& 0 \\
0 & \nu_2& . & . & .& 0 \\
. & . & .& .& .& . \\
. & . & .& .& .& . \\
. & . & .& .& .& . \\
0 & . & .& .& 0& \nu_n
\end{array} \right)\,,
\ee

\ni with $k=2n+r$. The parameters $\nu_1,\ldots, \nu_n$ characterize the
extension $\tilde{R}^k$, and thus they are the cohomology parameters. In
all physical situations, the subspace $R^{2n}$ of $R^k$ is associated with
the phase-space of a physical system, which possesses other symmetries than
those of $R^k$, like, for instance, rotations in each one of the subspaces
$R^n$ of $R^{2n}$. By requiring isotropy will fix these parameters to
coincide, $\nu_i=\nu,\forall i=1,\ldots,n$. For this case, the 2-cocycle
can be written as:
\be
\xi(\q_1,\p_1,\a_1;\q_2,\p_2,\a_2) =
\medio \nu (\q_2{\cdot}\p_1-\q_1{\cdot}\p_2)\,,
\ee

\ni where $\q_i$ are $n$-dimensional vectors corresponding to the first
$n$ coordinates (in the new basis), $\p_i$ correspond to the following $n$
coordinates,
and $\a_i$ to the remaining $r$ coordinates. Note that $\xi$ does not depend on
the $\a_i$, so the group $\tilde{R}^k$ can be written as H-W$_n\times R^r$,
where H-W$_n$ is the well-known Heisenberg-Weyl group. The group law for
$\tilde{R}^k$ can be written, in these new coordinates, as:
\begin{eqnarray}
\q\,'' & = & \q\,' + \q \nn \\
\p\,'' & = & \p\,' + \p  \\
\a\,'' & = & \a\,' + \a \nn \\
\z'' & = & \z'\z e^{\frac{i}{2} \nu (\p\,'\q - \q\,'\p)} \nn \,.
\end{eqnarray}

From the group law we see that if $\q$ is interpreted as coordinates, and
$\p$ as momenta, then $\nu=\hbar^{-1}$. Therefore, the cohomology
parameter for the (isotropic) Heisenberg-Weyl group can be identified with
$\hbar$. The variables $\a$ do not play any role, and can be factorized, as
we will see later.

Left and right invariant (under $\tilde{R}^k$) vector fields are:
\begin{equation}
\begin{array}{rcl}
\XL_{\q} & = & \parcial{\q} + \frac{\p}{2\hbar}  \Xo \\
\XL_{\p} & = & \parcial{\p} - \frac{\q}{2\hbar}  \Xo \\
\XL_{\a} & = & \parcial{\a}  \\
\end{array}\qquad
\begin{array}{rcl}
\XR_{\q} & = & \parcial{\q} - \frac{\p}{2\hbar}  \Xo \\
\XR_{\p} & = & \parcial{\p} + \frac{\q}{2\hbar}  \Xo \\
\XR_{\a} & = & \parcial{\a} \,, \\
\end{array}\,
\end{equation}

\ni and $\Xo=\parcial{\phi}$ is the vertical (left and right invariant)
vector field. The commutation relations for these vector fields are:
\be
[\XL_{q^i},\XL_{p^j}] = -\frac{1}{\hbar} \Xo\,,
\ee

\ni the rest of them being zero. In this way we reproduce the standard Weyl
commutation relations. Left and right invariant 1-forms for $R^k$ are
simply $d\x$, for $\x=\q,\p$ and $\a$. The quantization 1-form $\Theta$,
which for convenience we redefine with a factor $\hbar$, is:
\be
\Theta = \hbar\frac{d\z}{i\z} + \frac{1}{2}(\q{\cdot} d\p - \p{\cdot} d\q)\,.
\ee

Note that $d\Theta = d\q\wedge d\p$ is a pre-symplectic form on
$R^k$, with kernel the subspace $R^r$ spanned by the vectors $\a$. On the
quotient $R^k/R^r = R^{2n}$, it is a true symplectic form.
In fact, a partial complex structure $J$ can be introduced, of the form
$J= dp^i \otimes \XL_{q^i} - dq^i \otimes \XL_{p^i}$. $J$
turns to be a complex structure on the reduced space $R^k/R^r$.

The characteristic subalgebra, i.e. $Ker \Theta\cap\ Ker d\Theta$, is therefore
${\cal G}_{\Theta} = <\XL_{\a}>$, and the possible horizontal polarizations
we can find are of the form:
\be
{\cal P} = < \XL_{\a}, \alpha_i \XL_{q^i} + \beta_i\XL_{p_i},
            i=1,\ldots,n> \,,
\ee

\ni with restrictions on the real coefficients $\alpha_i,\beta_i$ such that it
is maximal, and horizontal with respect to $\Theta$. These restrictions
imply that the polarizations are full and symplectic. There are two
selected polarizations, ${\cal P}_p = <\XL_{\a}, \XL_{\q}>$ and ${\cal P}_q
= <\XL_{\a}, \XL_{\p}>$ leading to the representations in momentum and
configuration space, respectively. It should be stressed that all this
polarizations lead to equivalent representations of $R^k$, and in
particular the unitary operator relating the representations obtained with
${\cal P}_p$ and ${\cal P}_q$ is the Fourier transform. This is an outer
isomorphism of H-W$_n$, but it is inner in the Weyl-Symplectic group
$WSp(2n,R)$ \cite{Wolf}.

 Taking advantage of the natural complex structure of $R^{2n}\approx C^n$
(the one induced by $J$), we can allow for a complex polarization of the form:
\be
{\cal P}_c = <\XL_{\a}, \XL_{\q} + i\mu\XL_{\p}>\,,
\ee

 \ni where $\mu$ is a constant with the appropriate dimensions (from the
physical point of view, it will be a mass times a frequency, which makes
this polarization appropriate for the description of the Harmonic
Oscillator, see Sec. 3). This polarization leads to a representation in
terms of holomorphic (or anti-holomorphic) functions on $C^n$. It is
unitarily equivalent to the other representations, the unitary
transformation which relates it with the representation in configuration
space being the Bargmann transform. This is also an outer automorphism of
H-W$_n$, but it is inner in a certain subsemigroup of $Sp(2n,C)$
\cite{Wolf}.

Let us compute, for instance, the representation obtained with the
polarization ${\cal P}_q$. The equations $\XL_{\a} \Psi = 0$ leads to
wave functions not depending on the $\a$ variables (they trivially
factorize), and the equations $\XL_{p}\Psi =0$ lead to (together with
the equivariance condition $\Xo\Psi = i\Psi$):
\be
\Psi = \z e^{\frac{i}{2\hbar} \q{\cdot}\p} \Phi(\q\,)\,,
\ee

\ni where $\Phi(\q)$ is an arbitrary function of $\q$ (apart from
normalizability considerations). If we compute the action of the
right-invariant vector fields on these wave functions, we obtain:
\begin{eqnarray}
\XR_{\q} \Psi &=& \z e^{\frac{i}{2\hbar} \q{\cdot}\p}
               \left(\parcial{\q}\Phi(\q)\right)\\
\XR_{\p} \Psi &=& \z e^{\frac{i}{2\hbar} \q{\cdot}\p}
               \left(-\frac{i}{\hbar}\q\Phi(\q)\right)\,.
\end{eqnarray}

\ni This representation is unitarily
equivalent to the Schr\"{o}dinger representation (for each value of the
cohomology parameter $\hbar$), with respect to the measure 
$\mu=dq^1\wedge \cdots \wedge dq^n$, which in this case is invariant under
the group \Gt. This measure can be obtained contracting the left Haar measure
$\Omega^L$ with respect to all the vector fields in the polarization.

\vskip 0.4 cm

\subsection{The semisimple group $SU(2)$}
\label{SU(2)}

\vskip 0.3cm

Let us consider now an example, which in a certain sense is on the other
extreme to that
of the abelian group $R^k$. It is the semisimple group $SU(2)$, which has
trivial cohomology group $H^2(SU(2),U(1))=\{0\}$. For this reason all
2-cocycles on $SU(2)$ are coboundaries, and they will be classified
according to pseudocohomology classes only.

Making use of the realization of $SU(2)$ as $2\times 2$ complex matrices of
the form:
\be
\left(\begin{array}{cc}
z_1 & -z_2^* \\
z_2 & z_1^*
\end{array} \right)\,,
\ee

\ni with $|z_1|^2 + |z_2|^2 =1$, with matrix multiplication as group law, we
introduce stereographic projection
coordinates
$\{\eta\equiv \frac{z_1}{|z_1|},\,c\equiv \frac{z_2}{z_1},\,c^*=(c)^*\}$,
which are defined for $z_1\neq 0$. For $z_1=0$ another chart is needed, but
we shall forget about it and make use of the well-know geometrical properties
of the sphere to obtain the relevant results.

The group law in these coordinates is written as:
\begin{eqnarray}
\eta'' &=& \frac{\eta'\eta - \eta'{}^*\eta c^*{}'c}
            {\sqrt{ (1-\eta'{}^2c'c^*)(1-\eta'{}^{*2}c^*{}'c)}} \nn \\
c'' & = & \frac{c'\eta'{}^2 + c}{\eta'{}^2 - c^*{}'c} \\
c^*{}'' &=& (c'')^* \nn\,.
\end{eqnarray}

This set of coordinates has been chosen to make explicit the fibre bundle
structure of $SU(2)$ over the sphere,
$U(1)\rightarrow SU(2) \rightarrow S^2$, where $\eta\in U(1)$ is the
parameter of the fibre and the bundle projection is
$\pi(\eta,c,c^*)=(c,c^*)$, with $c,c^*$ the stereographic projection
coordinates of the sphere on the complex plane. The fibre bundle structure
is with respect to the right action of $SU(2)$,
$(\eta',c',c^*{}')*(\eta,0,0) = (\eta'\eta,c',c^*{}')$.

 As we mentioned before, due to the trivial cohomology of $SU(2)$, the only
2-cocycles on it are coboundaries. They are classified according to
pseudocohomology classes, which are in one-to-one correspondence with
integral co-adjoint orbits of $SU(2)$.

If we write $\eta=e^{i\varphi}$, then we can choose as representatives for the
pseudo-cohomology classes $\xi_{2j}(g',g) = 2j(\varphi'' - \varphi' - \varphi)$.
Using this pseudo-cocycle, we can introduce the following group law in the
direct product $SU(2)\times U(1)$:
\be g'' = g'*g, \qquad\qquad \z''= \z' \z e^{i2j (\varphi''-\varphi'-\varphi)}\,.
\ee

It is interesting to note that this is well-defined and therefore a group
law, or, in other words, $\xi_{2j}$ satisfies the 2-cocycle properties 
only if $2j \in Z$ (see
(\ref{cociclo})), from which we obtain the correct quantization condition of 
$j$ from the beginning, using only pseudo-cohomology considerations.

 Left-invariant vector fields are given by:
\begin{eqnarray}
\XL_{\eta} &=& \parcial{\varphi} \nn \\
\XL_{c} &=& \eta^{-2}\left[(1+|c|^2)\parcial{c} +
    \imedio c^*\left( \parcial{\varphi} + 2j\Xo\right)\right] \\
\XL_{c^*} &=& \eta^2\left[(1+|c|^2)\parcial{c^*} -
    \imedio c\left( \parcial{\varphi} + 2j\Xo\right)\right]
\nn\,,
\end{eqnarray}

\ni and right-invariant ones by:
\begin{eqnarray}
\XR_{\eta} &=& \parcial{\varphi} -2ic\parcial{c} + 2ic^*\parcial{c^*} \nn\\
\XR_{c} &=& \parcial{c} + c^*{}^2\parcial{c^*} -
          \imedio c^*\left(\parcial{\varphi} + 2j \Xo\right) \\
\XR_{c^*} &=& \parcial{c^*} + c^2\parcial{c} +
          \imedio c\left(\parcial{\varphi} + 2j \Xo\right)
\nn\,.
\end{eqnarray}

The commutation relations that these vector fields satisfy are:
\begin{eqnarray}
\left[ \XL_{\eta}, \XL_{c} \right] &=& -2i \XL_{c} \nn \\
\left[ \XL_{\eta}, \XL_{c^*} \right] &=& 2i \XL_{c^*} \\
\left[ \XL_{c}, \XL_{c^*} \right] &=& -i(\XL_{\eta} + 2j \Xo) \,. \nn
\end{eqnarray}

The left-invariant 1-forms, which are dual to the $\XL_{g^i}$'s, are:
\begin{eqnarray}
\theta^L{}^{\eta} &=& \frac{d\eta}{i\eta} - \medio \frac{ic^*}{1+|c|^2}dc
                         + \medio \frac{ic}{1+|c|^2} dc^* \nn\\
\theta^L{}^{c} &=& \frac{\eta^2}{1+|c|^2} dc \\
\theta^L{}^{c^*} &=& \frac{\eta^{-2}}{1+|c|^2} dc^* \nn \,,
\end{eqnarray}

\ni and the quantization 1-form is:
\be
\Theta = \frac{d\z}{i\z} - ij \frac{cdc^* - c^*dc}{1+|c|^2} \,.
\ee

The partial complex structure $J$ is given by $J=\theta^L{}^{c^*}\otimes
\XL_{c} - \theta^L{}^{c}\otimes \XL_{c^*}$.
Note that $d\Theta = 2ij\frac{dc\wedge dc^*}{(1+|c|^2)^2}$ is a
pre-symplectic form on $SU(2)$ which is projectable to $S^2$ defining the
standard symplectic structure on the sphere. The characteristic subalgebra
is $\GC=<\XL_{\eta}>$, if $j\neq 0$ and the whole $SU(2)$ algebra if $j=0$.
We shall not consider this last case, since it corresponds to the trivial
representation of $SU(2)$ (with spin zero), the wave functions being
constants on the group and the operators being zero. Therefore, let us
suppose $j\neq 0$. Then there are two possible polarizations, both full and
symplectic:
\begin{eqnarray}
{\cal P}_{c} &=& <\XL_{\eta}, \XL_{c^*} > \nn \\
{\cal P}_{c^*} &=& <\XL_{\eta}, \XL_{c} >\,.
\end{eqnarray}

These two polarizations lead to equivalent representations, since there is
an inner automorphism which takes one into the other. The first leads to a holomorphic
representation, while the second one leads to an anti-holomorphic one,
and the isomorphism is given by $(\eta,c,c^*) \rightarrow (\eta^*,c^*,c)$.

 Thus, we shall solve only the holomorphic polarization ${\cal P}_c$, with
solutions
\be
\Psi = \z (1+|c|^2)^{-j} \Phi(c)\,.
\ee

 If we had not obtained the integrallity condition $2j\in Z$ by
pseudo-cohomology considerations (if we had used a non-horizontal
polarization in the direct product $SU(2)\times U(1)$ with quantization
1-form $\Theta=\frac{d\z}{i\z}$), we would obtain it by geometrical
considerations,
i.e through the chart compatibility on the sphere. Making use of it, we
also obtain that the function $\Phi(c)$ is of the form:
\be
\Phi(c) = \sum_{l=0}^{2j} A_l c^l \,,
\ee

\ni with  arbitrary coefficients $A_l$, from which we recover the result that
the dimension of the representation labelled by $j$ is $2j+1$. Note that the
chart compatibility is what prevents the appearance of non-trivial null states
in going from the Lie algebra to the Lie group level, and this is also true
for infinite-dimensional representations of non-compact Lie groups, such as
$SL(2,R)$ (see \cite{Lang}, for instance), and even for infinite-dimensional Lie groups such as the Virasoro group \cite{Virasoro}.

The representation is given in terms of the action of the right-invariant
vector fields (which preserve the Hilbert space of polarized
wave functions):
\begin{eqnarray}
\XR_{\eta} \Psi &=& \z(1+|c|^2)^{-j} \left(-2ic\dot{\Phi}\right) \nn \\
\XR_{c} \Psi &=& \z(1+|c|^2)^{-j} \dot{\Phi}  \\
\XR_{c^*} \Psi &=& \z(1+|c|^2)^{-j} \left(c^2\dot{\Phi}-2jc\Phi\right)\,, \nn
\end{eqnarray}

\ni where $\dot{\Phi}=d \Phi(c)/dc$. $\XR_{\eta}$ is a diagonal
operator (generator of the Cartan subalgebra), $\XR_{c}$ can be interpreted
as an annihilation operator and $\XR_{c^*}$ as a creation operator. Since
the representation is finite-dimensional, there are maximal and minimal
weight states, given by $\Phi=c^j$ and $\Phi=1$, respectively. Note that we 
reproduce the standard commutation relations for $SU(2)$ in terms of 
creation and annihilation operators with the definitions:
\begin{eqnarray}
\hat{J}_0 &\equiv& \imedio (\XR_{\eta} + 2j\Xo) \nn \\
\hat{J}_+ &\equiv& \frac{i}{\sqrt{2}} \XR_{c^*}  \\
\hat{J}_- &\equiv& \frac{i}{\sqrt{2}} \XR_{c}  \nn 
\end{eqnarray}

\ni This implies, in particular, that the generator $\XR_{\eta}$
should be redefined by $\XR_{\eta}{}'=\XR_{\eta}+2j\Xo$, in accordance with 
the general theory, since in this case $\lambda^0_{\eta}=2j,\,
\lambda^0_{c}=\lambda^0_{c^*}=0$. 

There is an invariant measure of the form 
$\mu=\theta^{Lc}\wedge \theta^{Lc^*}= \frac{dc\wedge dc^*}{(1+|c|^2)^2}$, which
is obtained by contracting the left Haar measure $\Omega^L$ with $\XL_{\eta}$.
With respect to this measure $\hat{J}_0$ is hermitian, and $\hat{J}_+$ and 
$\hat{J}_-$ are the adjoint of each other.

\section{Algebraic Anomalies}\label{Algebraic Anomalies}

In Sec.2, we introduced  the concept of full and symplectic polarization
subalgebra intended to reduce the representation obtained through the
right-invariant vector fields acting on equivariant functions on the group.
It contains ``half'' of the symplectic vector fields as well as the entire
characteristic subalgebra. If the full reduction is achieved, the whole set
of physical operators can be rewritten in terms of the basic ones, i.e.
those which are the right version of the left-invariant generators in
$J{\cal P}\oplus J^2{\cal P}$. For instance, the energy operator for the
free particle can be written as $\frac{\hat{p}^2}{2m}$, the angular
momentum in 3+1 dimensions is the vector product ${\bf \hat{x}}\times {\bf
\hat{p}}$, or the energy for the harmonic oscillator is
$\hat{c}^{\dag}\hat{c}$ (note that, since we are using first-order
polarizations, all this operators are really written as first-order
differential operators, and, for instance, the energy operator in momentum
space is written as $\hat{E}\Psi=\frac{p^2}{2m}\Psi$, which is a
zeroth-order differential operator, indeed).

However, the existence of a full and symplectic polarization is guaranteed
only for semisimple and solvable groups \cite{Kirillov}. We define  an {\bf
anomalous} group \cite{Anomalias} (see also \cite{ConfDubna,Bregenz}) as a 
group $G$ which, for some central
extension \Gt characterized by certain values of the (pseudo-)cohomology
parameters, does not admit any polarization which is full and symplectic.
These values of the (pseudo-)cohomology parameters are called the {\bf
classical} values of the anomaly, because they are associated with some
coadjoint orbits of the group \Gt (generally exceptional orbits, of lower
dimension),
that is, with the classical phase space of some physical system (see the
discussion in Sec. 2 on the relation between (pseudo)-cohomology parameters
and coadjoint orbits of the group \Gt).

Anomalous groups feature another set of values of the (pseudo-)cohomology
parameters, called the {\bf quantum} values of the anomaly, for which the
carrier space associated with a full and symplectic polarization contains
an invariant subspace. For the classical values of the anomaly, the
classical solution manifold undergoes a reduction in dimension thus
increasing the number of (non-linear) relationships among Noether
invariants (invariant relations which characterize the lower dimensional 
exceptional orbits. These can be defined as a set of
equations of the form $f_i(F_{g^j})=0\,,i=1,\ldots,k$, where the functions
$f_i:\tilde{\cal G}^*\rightarrow R$ are in involution  and satisfy
$\{f_i,F_{g^i}\}=\gamma_{ij}^kf_k$ with $\gamma_{ij}^k$ functions on
$\tilde{\cal G}$. See \cite{Levi-Civita}
for a discussion on invariant relations in the context of Rational Mechanics),
whereas for the quantum values the number of basic operators decreases on
the invariant (reduced) subspace due to the appearance of (higher-order)
relations among the quantum operators ("quantum invariant relations", which 
can be defined as a set of equations of the form
$A_i\Psi=0\,,i=1,\ldots,k$, where $A_i\in {\cal U}\tilde{\cal G}^L$
close an algebra and satisfy $[A_i,\XL_{g^j}] = B_{ij}^k A_k$, with
$B_{ij}^k \in {\cal U}\tilde{\cal G}^L$).
The anomaly lies in the fact that
the classical and quantum values of the anomalies do not coincide, but
there is a "shift" between them or even there is no relation at all among
them. The reason is that, when passing from the classical invariant relations
to the quantum ones, problems of normal ordering can appear which
"deform" the classical Poisson algebra between the functions defining the
classical invariant relations.

We must remark that the anomalies we are dealing with in this paper are
of {\it algebraic} character in the sense that they appear at the Lie algebra
level,
and must be distinguished from the {\it topologic anomalies} which are
associated with the non-trivial homotopy of the (reduced) phase space
\cite{Frachall}.

The non-existence of a full and/or symplectic polarization for certain
values of the (pseudo-)cohomology parameters (the classical values of the
anomaly) is traced back to the presence in the characteristic subalgebra of
some elements the adjoint action of which are not diagonalizable in the
complementary subspace of ${\cal G}_C$ in $\tilde{\cal G}$. In other words, no maximal isotropic
subspace for the symplectic 2-form on the coadjoint orbit is a Lie
subalgebra. The anomaly problem here presented parallels that of the
non-existence of invariant polarizations in the Kirillov-Kostant co-adjoint
orbits method \cite{Gotay,Kirillov}, and the conventional anomaly problem
in Quantum Field Theory which manifests itself through the appearance of
central charges in the quantum current algebra, absent from the classical
(Poisson bracket) algebra \cite{Jackiw}.

The full reduction of  representations in anomalous cases will be achieved
by means of a generalized concept of (higher-order) polarization (see Sec. 5).
Higher-order polarizations are needed to accommodate the "quantum invariant
relations" inside the polarization, and these are given, in general, by
operators in the (left) enveloping algebra.

Let us consider a couple of anomalous groups, one of them finite
dimensional (the Schr\"{o}dinger group) and the other one infinite dimensional
(the Virasoro group).

\vskip 0.4cm

\ni {\large \it The Schr\"{o}dinger group}

\vskip 0.3cm

To illustrate the Lie algebra structure of an anomalous group, let us first
consider the example of the Schr\"{o}dinger group. This group,  or rather  the
non-extended $n$-dimensional version of it, was considered in Ref. \cite{Kirillov} as an
example of a group not possessing an {\it admissible} subalgebra (the
equivalent to a full and symplectic polarization in our context). In the
simplest 1-dimensional case, $G$ is the semidirect action of the symplectic
group $Sp(1,R)\approx SL(2,R)$ on the phase space $R^2$. We will consider a
central extension of it, the Schr\"odinger group \Gt, which is given by the 
semidirect action of
$SL(2,R)$ on the Heisenberg-Weyl group. This group includes as subgroups
the symmetry group of the free particle, the Galilei Group, as well as the
symmetry group of the ordinary harmonic oscillator and the ``repulsive''
harmonic oscillator (with imaginary frequency), usually known as Newton
groups \cite{Niederer}. From the mathematical point of view, it can be
obtained from the Galilei (or from either of the Newton) groups by
replacing the time subgroup with the three-parameter group $SL(2,R)$. In
fact, those kinematical subgroups are associated with different choices of
a Hamiltonian inside $SL(2,R)$.

Let us parameterize the Schr\"{o}dinger group by $(x,v,a,t,c,\z)$, where
$(x,v,\z)$ parameterize the Heisenberg-Weyl subgroup ($\z\in U(1)$), and
$(a,t,c)$ are the parameters for the $SL(2,R)$ subgroup, with $a\in R-\{0\}$,
for which we use the following Gauss decomposition \cite{Anomalias}:
\begin{equation}
\left(\begin{array}{cc}
\alpha & \beta \\
\gamma & \delta
\end{array}\right) =
\left(\begin{array}{cc}
1& t \\ 0 & 1
\end{array}\right)
\left(\begin{array}{cc}
1 & 0 \\ c & 1
\end{array}\right)
\left(\begin{array}{cc}
a^{-1} & 0 \\ 0 & a
\end{array}\right)
\end{equation}

\ni with $\alpha\delta-\beta\gamma=1$. For our purposes we only need the Lie
algebra (see \cite{Anomalias} for a
detailed account of the group law and the expressions of vector fields),
which is given (in terms of left-invariant vector fields) by:
\begin{equation}\begin{array}{rcl}
\[\XL_a,\XL_t\]&=&2\XL_t \\
 \[\XL_t,\XL_c\]&=&\XL_a\\ \[\XL_a,\XL_c\]&=&-2\XL_c \\ \[\XL_t,\XL_x\]&=&0 \\
\[\XL_t,\XL_v\]&=&-\XL_x
  \end{array} \;\;\;\; \begin{array}{rcl}\[\XL_a,\XL_x\]&=&\XL_x \\
\[\XL_a,\XL_v\]&=&-\XL_v \\ \[\XL_c,\XL_x\]&=&-\XL_v
 \\ \[\XL_c,\XL_v\]&=&0 \\ \[\XL_x,\XL_v\]&=& m\XL_0\,, \label{algebra}
\end{array}
\end{equation}

Analysing the Lie-algebra 2-cocycle $\Sigma$ (see Sec. 2.2), we deduce that this central
extension is associated with the (exceptional) 2-dimensional orbit of the
Schr\"{o}dinger group, since it contains the entire $SL(2,R)$ subalgebra
$<\XL_t,\XL_a,\XL_c>$ (the characteristic subalgebra, or isotropy subalgebra
of the coadjoint orbit) in its kernel. According to the general scheme, the
characteristic subalgebra should enter
any full and symplectic polarization but such a polarization does not
exist, and this can be traced back to the fact that the partial complex
structure $J=\theta^L{}^{v}\otimes \XL_{x} -
\theta^L{}^{x}\otimes\XL_{v}$ is not preserved by its kernel (the $SL(2,R)$
subalgebra), implying that we cannot project it onto a complex structure on
the classical phase-space. We can only find a symplectic, non-full
polarization,
\begin{equation}
{\cal P}=<\XL_t,\XL_a,\XL_x>\,, \label{non-full}
\end{equation}

\ni and a full, but non-symplectic one,
\begin{equation}
{\cal P}_C=<\XL_t,\XL_a,\XL_c>\,\approx \,SL(2,R)\,. \label{non-symplectic}
\end{equation}

\ni Quantizing with the non-full polarization (\ref{non-full}) results in
a breakdown of the naively expected correspondence between the operators
$\XR_t,\XR_a,\XR_c$ and the basic ones $\XR_x,\XR_v$, i.e. the one suggested by
the Noether invariants (see Sec. 2) which can be written as \cite{Anomalias}:
\begin{eqnarray}
F_t&=&-\frac{1}{2m}F_x^2  \nn \\
F_a&=&-\frac{1}{m}F_xF_v \label{invariant-relations}\\
F_c&=& \frac{1}{2m}F_v^2 \nn \,.
\end{eqnarray}

\ni This relations characterize the two-dimensional (exceptional) coadjoint
orbit, being the {\it invariant relations} mentioned above.

On the other hand, quantizing with the non-symplectic polarization
(\ref{non-symplectic})
leads to an unconventional representation in which the wave functions
depend on both $x$ and $p$ variables, which contains two irreducible
components (see \cite{Hove}) distinguished by the eigenvalues of the
parity operator, which is not in the group. This particular example shows
the principal drawback of the higher-order polarization technique,
since there are no means of obtaining the parity operator from the group nor
its enveloping algebra. As happened with the other polarization, the
operators  $\XR_t,\XR_a,\XR_c$, neither, are expressed in terms of
$\XR_p,\XR_x$.

In both cases the operators $\XR_t,\XR_c$ behave as if they also were basic
operators, i.e. as if
\begin{equation}
\[\XL_t,\XL_c\]=\XL_a+2k\XL_0 \label{deformado}
\end{equation}

\ni would replace the corresponding commutator in (\ref{algebra}) with a
non-trivial value of the anomaly
(pseudo-extension parameter) $k$. In other words, both quantizations seem
to correspond to a four-dimensional (of maximal dimension) orbit, for which
there are no "quantum invariant relations" between operators (the quantum
counterpart of the invariant relations between Noether invariants).

Therefore, we should start with the Lie algebra (\ref{algebra}) with
"deformed" commutators like in (\ref{deformado}). In this way, we are
considering a whole family (this kind of pseudo-extension
corresponds to a family of four dimensional orbits associated with the 1-sheet
hyperboloid of $SL(2,R)$. To consider the rest of the coadjoint orbits,
associated with the 2-sheet hyperboloid or the cone, a different
pseudo-extension is required, associated with the compact Cartan subgroup of
$SL(2,R)$) of coadjoint orbits, all of them
four-dimensional, except for $k=0$ which is two-dimensional. For all of these
pseudoextensions, the non-full polarization (\ref{non-full}) is now full
and symplectic, since the characteristic subalgebra is smaller
(corresponding to the fact that for the four-dimensional orbits the
isotropy subalgebra is smaller), ${\cal G}_C=<\XL_a>$. In the next
section we shall see that the fact that there is a full and symplectic
polarization does not guarantee the irreducibility of the representation,
and that there exist a value of the parameter $k$ for which the
representation obtained is reducible, admitting an invariant subspace.

\vskip 0.4cm

\ni {\large \it The Virasoro group}

\vskip 0.3cm

Let us comment very briefly on the relevant, although less intuitive,
example of the infinite-dimensional Virasoro group. Its Lie algebra can
be written as
\begin{equation}
\left[\XL_{l_n},\XL_{l_m}\right]=-i(n-m)\XL_{l_{n+m}}-\frac{i}{12}(cn^3-c'n)\Xi\,,
\end{equation}

\ni where $c$ parameterizes the central extensions and $c'$ the
pseudo-extensions. As is well known, for the particular case in which
$\frac{c'}{c}=r^2\,,r\in N, r>1$, the co-adjoint orbits admit no invariant
K\"{a}hlerian structure.
In the present approach, this case shows up as an algebraic anomaly. In
fact, the characteristic subalgebra is given by
$\GC=<\XL_{l_0},\XL_{l_{-r}},\XL_{l_{+r}}>$, which is not fully
contained in the non-full (but symplectic) polarization
${\cal P}^{(r)}=<\XL_{l_{n\leq 0}}>$. There also exists a full
polarization ${\cal P}_C=<\XL_{l_{kr}}>\,,r>1, k=-1,0,1,2,3,...$ which
is not symplectic since none of the symplectic generators
with labels $l_{{\pm} r'}, r'\neq kr$ are included in the polarization.
A detailed description of the representations of the Virasoro group can be
found in \cite{Virasoro} and references therein.

\section{Higher-order Polarizations}

In general, to tackle  situations like those mentioned above, it is necessary
to generalize the notion of polarization. Let us consider the universal enveloping
algebra of left-invariant vector fields,
${\cal U}\tilde{\cal G}^L$.
We define a {\bf higher-order polarization} ${\cal P}^{HO}$ as a maximal
subalgebra of ${\cal U}\tilde{\cal G}^L$ with no intersection with the
abelian subalgebra of powers of $\Xo$. With this definition a higher-order
polarization contains the maximum number of conditions (expressed in terms
of differential operators) compatible
with the equivariance condition of the wave functions and with the action
of the physical operators (right-invariant vector fields).

We notice that now the vector space of functions annihilated by a
higher-order polarization is not, in general, a ring of
functions and therefore there is no corresponding foliation; that is,
they cannot be characterized by saying that they are constant along
submanifolds. If this were the case, it would mean that the
higher-order polarization was the enveloping algebra of a first-order
polarization and, accordingly, we could consider the submanifolds
associated with this polarization. In this sense the concept of
higher-order polarization generalizes that of first-order
polarization.

The definition of higher-order polarization given above is quite general.
In all studied examples higher-order polarizations adopt a more definite
structure closely related to given first-order (non-full and/or
non-symplectic) ones.
According to the until now studied cases, higher-order polarizations can be
given a more operative definition: {\it A higher-order polarization is a
maximal subalgebra of ${\cal U}\tilde{\cal G}^L$ the ``vector
field content'' of which is a first-order polarization}.
By ``vector field content'' of a subalgebra ${\cal A}$ of
${\cal U}\tilde{\cal G}^L$ we mean the following:
Let $V({\cal A})$ be the vector space of
complex functions on \Gt defined by
\begin{equation}
V({\cal A})=\{f\in {\cal F}_C(\tilde{G})\; /\; A{\cdot} f=0, \forall A\in {\cal A}\} \,.
\end{equation}

\ni Now we consider the ring $R({\cal A})$ generated by elements
of $V({\cal A})$. With $R({\cal A})$ we associate the set of left-invariant
vector fields defined by
\begin{equation}
L_{\tilde{X}^L} h=0\,, \;\;  \forall h\in R({\cal A})\,.
\end{equation}

\ni This set of left-invariant vector fields is a Lie subalgebra of
$\tilde{{\cal G}}^L$ and defines the vector field content of ${\cal A}$, which
proves to be a first-order polarization.

 Even though a higher-order polarization contains the maximal number
of conditions (expressible in terms of differential equations) there can be
non-trivial operators acting on the Hilbert space of wave functions
which are not differential, and therefore
are not contained in the enveloping algebra. This implies that higher-order
polarizations will not guarantee, in general, the irreducibility of the
resulting representation, since there can be non-trivial and non-differential
operators acting on the resulting Hilbert space and commuting with the
representation.

A simple example suggesting the need of a generalization of the concept of
higher-order polarization corresponds to the non-irreducible representation
associated with the non-symplectic polarization (\ref{non-symplectic}) of
the Schr\"{o}dinger group. This representation cannot be further reduced by
enlarging (\ref{non-symplectic}) to a higher-order polarization ${\cal
P}^{HO}$. A full reduction requires the inclusion in ${\cal P}^{HO}$ of the
parity operator commuting with the representation. The generalization of
the concept of higher-order polarization so as to include this kind of
operators not reachable by the enveloping algebra of the group, as well as
a constructive characterization of those operators deserves a separate
study. We outline a possible way to deal with the problem.
We consider the group of unitary
automorphisms $A$ of the space ${\cal F}$ of complex valued functions on \Gt satisfying the
equivariance condition (\ref{equivariance}) (or, rather, of the completion of 
it in the scalar product defined by the left invariant Haar measure). This 
(huge) group $A$ will replace  the role of the enveloping algebra, and the 
role of the higher-order polarization will be played by a {\it polarizing 
subgroup} $P$ of $A$. This polarizing subgroup $P$ must be a maximal subgroup 
of $A$ satisfying the following conditions:
\begin{itemize}
\item[i)] $P\cap U(1)=\{I\}$, where $U(1)$ is the vertical subgroup.
\item[ii)] If $T\subset A$ is the left regular representation of \Gt 
restricted to the space of equivariant functions, then $T$ must be in the 
normalizer of $P$, i.e. $[P,T]\subset P$.
\end{itemize}

The reduced space ${\cal H}$ is defined as the set of equivariant functions 
on which all the elements of $P$ act as the identity. Then $T$, when 
restricted to ${\cal H}$, is irreducible. To see it, suppose that $B$ is an 
invariant subspace of
${\cal H}$, and $p$ the (self-adjoint) projector on it, which commutes
with $T$. Then $e^{i\lambda p}=I+(e^{i\lambda}-1)p$ is a unitary operator
on ${\cal H}$ commuting with $T$. It can be trivially extended to a
unitary operator on ${\cal F}$ commuting with $T$, and
therefore, since $P$ is maximal, it must be contained in $P$. But $P$ acts
trivially on ${\cal H}$, and therefore $p$ annihilates ${\cal H}$. Thus,
$B$ is trivial and $T$ is irreducible.

Let $P'$ be the minimal subgroup of $P$ determining the same Hilbert
space ${\cal H}$ satisfying the condition $P'\Psi=\Psi$. Let $P_0'$ be the
connected component of $P'$, and ${\cal P}'$ its Lie algebra, which,
according to the Stone-von Newmann theorem, will be constituted by self-adjoint
 operators in (some dense domain of) ${\cal F}$. Clearly, the condition 
${\cal P}'\Psi=0$ is equivalent to the condition $P'\Psi=\Psi$ and may 
determine the same Hilbert space ${\cal H}$
 only if $P'$ is connected, i.e. if $P'/P_0$ is trivial.

For the cases in which a first-order (full and symplectic) polarization ${\cal P}^1$ exists, and
it is enough to reduce completely the representation $T$, it is clear that ${\cal P}'$ will coincide
with ${\cal P}^1$ and will be constituted by first order differential
operators.


Also, for the cases in which a higher-order polarization ${\cal P}^{HO}$
is enough to obtain an irreducible representation, ${\cal P}'$ will coincide
with ${\cal P}^{HO}$, and will be constituted by higher-order differential
operators.


Those cases, like the example commented above of the Schr\"{o}dinger group
with the polarization (\ref{non-symplectic}), in which a higher-order
polarization is not enough to obtain an irreducible representation, lie in
the category of groups for which ${\cal P}'$ contains non-differential
self-adjoint operators or $P'$ is not connected (or both of them). The
example of the Schr\"{o}dinger group
with the polarization (\ref{non-symplectic}) lies in the second category,
since the operator we need to completely reduce the representation is the
parity operator, which is discrete, and therefore belongs to $P'/P_0'$.

To see how a higher-order polarization operates in practice, we shall
consider first a simple non-anomalous example like the Harmonic Oscillator
in configuration space, and later  we will come back to the cases of the
Schr\"{o}dinger and Virasoro groups.

\vskip 0.4cm

\ni {\large \it The Harmonic Oscillator}

\vskip 0.3cm

The (quantum) Harmonic Oscillator group (we shall restrict to the
one-dimensional case, since it presents all the interesting features and the
treatment is far simpler), as the Galilei group, is a semidirect product
of the time translations and the Heisenberg-Weyl group. The difference
relies precisely on the semidirect actions, which correspond to different
choices for uniparametric subgroups of $Sp(2,R)\approx SL(2,R)$ acting on
H-W as linear canonical transformations. For the case of the Galilei group,
the time translations are those generated by $\hat{P}^2$, and constitutes a
non-compact subgroup, while for the harmonic oscillator group, time
translations are generated by $\hat{P}^2 + \hat{X}^2$, which corresponds to
the compact subgroup $SO(2)$ of $SL(2,R)$.

The harmonic oscillator group possesses nontrivial group cohomology, but
the pseudo-cohomology is trivial (although pseudo-extensions can be
introduced, all of them lead to equivalent representations).

 The group law for the harmonic oscillator group can be obtained from this
semidirect action (in fact, it can be seen as a central extension of the
Euclidean group $E(2)$), see \cite{Anomalias}:
 \begin{eqnarray}
 t''&=&t'+t \nn \\
 x''&=&x+x'\cos\omega t+\frac{p'}{m\omega}\sin\omega t \nn \\
 p''&=&p+p'\cos\omega t-m\omega x'\sin\omega t    \label{NRGL}  \\
 \zeta''&=&\zeta'\zeta e^{\frac{i}{2\hbar}(x'p\cos\omega t-p'x\cos\omega t
          +(\frac{p'p}{m\omega}+\omega x'x)\sin\omega t )}  \nn \, .
 \end{eqnarray}

It is easy to see that under the limit $\omega\rightarrow 0$ (which
corresponds to a group contraction in the sense of In\"{o}n\"{u} and Wigner)
we obtain the group law for the Galilei group.

The left-invariant vector fields are:
 \begin{eqnarray}
 \tilde{X}^{L}_{t}&=&\frac{\partial}{\partial t}+\frac{p}{m}\frac{\partial}
     {\partial x}-m\omega^{2}x\frac{\partial}{\partial p} \nn \\
 \tilde{X}^{L}_{x}&=&\frac{\partial}{\partial x}-\frac{p}{2\hbar}\Xo \nn \\
 \tilde{X}^{L}_{p}&=&\frac{\partial}{\partial p}-\frac{x}{2\hbar}\Xo  \label{NRLF}  \\
 \tilde{X}^{L}_{\zeta}&=&\parcial{\phi}\equiv\Xo \, , \nn
 \end{eqnarray}

 \ni and the right ones are:
 \begin{eqnarray}
 \tilde{X}^{R}_{t}&=&\frac{\partial}{\partial t} \nn \\
 \tilde{X}^{R}_{x}&=&\cos\omega t\frac{\partial}{\partial x}-
  m\omega \sin\omega t \frac{\partial}{\partial p}+
      \frac{1}{2\hbar}(p\cos\omega t+m\omega x\sin\omega t)\Xo   \\
 \tilde{X}^{R}_{p}&=&\cos\omega t\frac{\partial}{\partial p}+\frac{1}{m\omega}
        \sin\omega t\frac{\partial}{\partial x}-\frac{1}{2\hbar}(x\cos\omega t-
    \frac{p}{m\omega}\sin\omega t)\Xo \nn \\
 \tilde{X}^{R}_{\zeta}&=&\parcial{\phi}\equiv\Xo \, . \nn
 \end{eqnarray}

 \ni The commutation relations for theses vector fields are:
 \begin{eqnarray}
 \left[\tilde{X}^{R}_{t}, \; \tilde{X}^{R}_{x}\right]&=&
        -m\omega^{2}\tilde{X}^{R}_{p} \nn \\
 \left[\tilde{X}^{R}_{t}, \; \tilde{X}^{R}_{p}\right]&=&
        \frac{1}{m}\tilde{X}^{R}_{x} \\
 \left[\tilde{X}^{R}_{x}, \; \tilde{X}^{R}_{p}\right]&=&
       -\frac{1}{\hbar}\Xo \,. \nn
 \end{eqnarray}

The quantization 1-form $\Theta$ (we redefine it with a factor $\hbar$) is:
\be
\Theta = \hbar\frac{d\z}{i\z} + \medio(pdx-xdp) - \left(\frac{p^2}{2m} +
            \medio m\omega^2 x^2\right) dt \,,
\ee

\ni and the characteristic subalgebra is $\GC = <\XL_{t}>$.
The partial complex structure is given by $J=\theta^L{}^{p}\otimes \XL_{x}
- \theta^L{}^{x}\otimes \XL_{p}$.
If we look for
first order polarizations, we find that we are not able to find a full and
symplectic real polarization (in some sense, the Harmonic Oscillator is
an anomalous system, see \cite{Streater}), only complex polarizations
(making use of the natural complex structure of the phase space $R^2\approx
C$ of the system, induced again by $J$) can be full and symplectic. They are
of the form:
\be
{\cal P}_{\pm} = <\XL_t, \XL_x {\pm} im\omega \XL_p>\,,
\ee

\ni and lead to the Bargmann-Fock representation of the harmonic oscillator
in terms of (anti-)holomorphic functions (see \cite{Wolf,Relativistic}).

 But now we could be interested in using only real polarizations for obtaining
the configuration (or momentum) space representation, and this can only be
achieved if we resort to  higher-order polarizations. For this simple case
it is an easy task to obtain the higher-order polarizations, since we only
have to consider the subalgebras of the left-enveloping algebra generated
by the Casimir
\be
\XL_t -\frac{i\hbar}{2m}\left(\XL_x\right)^2 -
       \frac{i\hbar m\omega}{2}\left(\XL_p\right)^2\,,
\ee

\ni and $\XL_p$ or $\XL_x$. The subalgebra generated by the Casimir and
$\XL_t$ is not maximal, since we can still add $\XL_x{\pm} im\omega\XL_p$.
Therefore, there are essentially two real higher-order polarizations,
${\cal P}^{HO}_x = <\XL_t -\frac{i\hbar}{2m}\left(\XL_x\right)^2, \XL_p>$,
which leads to the representation in configuration space, and
${\cal P}^{HO}_p = < \XL_t - \frac{i\hbar m\omega}{2}\left(\XL_p\right)^2,
\XL_x>$, leading to the representation in momentum space. These two
representations are unitarily equivalent, the unitary transformation being
the Fourier transform, and are also unitarily equivalent to the
Bargmann-Fock representation through the Bargmann transform (see the
comments on the case of the abelian group $R^k$).

Let us consider, for instance, the polarization ${\cal P}^{HO}_x$ leading
to configuration space. The solutions to the polarization equations are:
 \begin{eqnarray}
 \tilde{X}^{L}_{p}\Psi=0&\rightarrow &\Psi=\zeta
      e^{-\frac{i}{\hbar}\frac{px}{2}}\Phi(x,t) \label{EqXS}  \\
 (\tilde{X}^{L}_{t}-\frac{i\hbar}{2m}
     \tilde{X}^{L}_{x}\tilde{X}^{L}_{x})\Psi=0&\rightarrow &
       i\hbar\frac{\partial\Phi}{\partial t}=
        -\frac{\partial^{2}\Phi}{\partial x^{2}}+
            \frac{1}{2}m\omega^{2}x^{2}\Phi\, . \nn
 \end{eqnarray}

 The last equation is the well-know Schr\"{o}dinger equation for the
 harmonic oscillator in configuration space, with the standard solutions in
terms of Hermite
 Polynomials:
 \begin{eqnarray}
 \Phi\equiv\sum^{\infty}_{n=0}c_{n}\Phi_{n}(x,t)&=&\sqrt{\frac{\omega}{2\pi}}
        \left(\frac{m\omega}{\hbar\pi}\right)^{1/4} \nn\\
     & & \times   \sum^{\infty}_{n=0}\frac{c_{n}}{2^{n/2}\sqrt{n!}}
      e^{-\frac{m\omega}{2\hbar}x^{2}}e^{-i(n+1/2)\omega t}
       H_{n}(\sqrt{\frac{m\omega}{\hbar}}x)\, .
 \end{eqnarray}

The scalar product can be obtained from the (left-invariant) Haar measure 
on the group. Indeed, $i_{\XL_p} \Omega^L=dx\wedge dt$ is an invariant 
measure on the quotient space $G/G_p$, with $G_p$ the subgroup generated
by $\XL_p$:
\begin{equation}
<\Psi'|\Psi> = \int dxdt \Phi'(x,t)^*\Phi(x,t) \, .
\end{equation}


We would like to stress that although the representations here obtained
in configuration space (or its analogous in momentum space) are unitarily
equivalent to the one obtained in the Bargmann-Fock space, the latest
requires of a process of "unitarization" to obtain the correct energy of the
vacuum $E_0 = 1/2 \hbar\omega$, which otherwise would be zero
(see \cite{Wolf}). In the literature this problem was solved recurring to
the ``metaplectic correction'' (see \cite{Woodhouse}), and here we obtain
the correct result resorting only to higher-order polarizations. Moreover,
this fact can be seen as a reminiscence of the anomaly of the Schr\"{o}dinger
group, which causes the correct ordering of the operators.

\vskip 0.4cm

\ni {\large \it The Schr\"{o}dinger group}

\vskip 0.3cm

Now we consider the case of the Schr\"{o}dinger group and the representation
associated
with the non-full polarization (\ref{non-full}) for which the operators
$\XR_{t},\XR_{c}$ are basic.
As stated before, for commutation relations like (\ref{deformado}), the
polarization (\ref{non-full}) becomes full and symplectic, as far as
$k\neq 0$. Thus one would think that the associated representations (with
two degrees of freedom) are irreducible.
However, for a particular value (the quantum value of the anomaly)
$k=\frac{1}{4}$, the representation of the Schr\"{o}dinger
group {\it becomes reducible} and,
{\it on the invariant subspace}, the operators $\XR_{t},\XR_{c}$ do
really express as  $\hat{p}^2/2m,\hat{x}^2/2$, respectively.
 The invariant subspace is constituted by the solutions
of a second-order polarization which exists only for $k=\frac{1}{4}$:
\begin{equation}
{\cal P}^{HO}=<\XL_t,\XL_{a},\XL_x, \XL_{c}-\frac{i}{2m}(\XL_{v})^2>\,,
\end{equation}

This result indicates that for the (four-dimensional) coadjoint orbit
associated with the value $k=\frac{1}{4}$ of the pseudo-extension parameter,
although no classical invariant relations like (\ref{invariant-relations})
exist, there exist  "quantum invariant relations", in such a way that the
quantum system possesses only one degree of freedom (see \cite{symplin} for
a detailed discussion on this question). The quantum invariant relations
are of the form:
\begin{eqnarray}
\XL_t{}^{HO} \Psi &=&0 \nn \\
\XL_a{}^{HO} \Psi &=&0 \label{quantuminvrel}\\
\XL_c{}^{HO} \Psi &=&0 \nn \,,
\end{eqnarray}

\ni where the second-order operators are given by
\begin{eqnarray}
\XL_t{}^{HO}  &=& \XL_t -\frac{i}{2m}(\XL_x)^2 \nn \\
\XL_a{}^{HO}  &=& \XL_a -\frac{i}{m}\XL_v\XL_x \\
\XL_c{}^{HO}  &=& \XL_c + \frac{i}{2m}(\XL_v)^2 \nn \,.
\end{eqnarray}

In fact, note that the polarization ${\cal P}^{HO}$ is equivalent to the
one given by:
\begin{equation}
<\XL_t{}^{HO},\XL_{a}{}^{HO},\XL_{c}{}^{HO},\XL_x>\,,
\label{HOPolarization}
\end{equation}

\ni and $k$ must be $k=\frac{1}{4}$ for this to be a higher-order
polarization, or, in other words, for (\ref{quantuminvrel}) to constitute
true "quantum invariant relations". The fact that classical and quantum
invariant relations are realized for different values of $k$ ($k=0$ and
$k=\frac{1}{4}$, respectively) can be thought of as being due to normal
ordering problems, in the operator $\XL_{a}{}^{HO}$ to be precise.

It is worth noting that in the polarization (\ref{HOPolarization}), the
characteristic subalgebra which, for $k=0$, is given by
$\GC=<\XL_t,\XL_a,\XL_c>\approx SL(2,R)$, has been substituted by a
higher-order characteristic subalgebra
$\GC^{HO}=<\XL_t{}^{HO},\XL_{a}{}^{HO},\XL_{c}{}^{HO}>$, which, for
$k=\frac{1}{4}$, satisfy identical commutation relations but commute with
the generators $\XL_x$ and $\XL_v$, and therefore can be included in a
polarization (although of higher-order type).

Physical applications of this particular representation,
although in the harmonic oscillator realization are found in Quantum Optics
\cite{Optica} although no reference to the connection between anomalies and
the restriction of $k$ has been made. Note that when restricted to the
$SL(2,R)$ subgroup, the representation obtained is reducible decomposing in
two irreducible ones with Bargmann indices $k=\frac{1}{4}$ and $k+\medio =
\frac{3}{4}$.

\vskip 0.4cm

\ni {\large \it The Virasoro group}

\vskip 0.3cm

In a similar way, in the case of the Virasoro group, for particular values
of the parameters $c,c'$ or equivalently $c,h\equiv \frac{c-c'}{24}$ given
by the Kac formula \cite{Kac}, the ``quantum values" of the anomaly,
the representations given by the first order non-full (symplectic)
polarizations are reducible since there exist invariant subspaces
characterized by certain higher-order polarization equations
\cite{Virasoro}, which constitute the "quantum invariant relations".
Note that there is no one-to-one correspondence
between the values of $c'/c$ characterizing the coadjoint orbits of the
Virasoro group (the classical values of the anomaly) and the values
allowed by the Kac formula (the quantum values of the anomaly), a fact
which must be interpreted as a breakdown of the notion of classical
limit.


\section{Comments and outlooks}

Let us comment further on the relationship between the present formalism
and the more conventional method formulated on the co-adjoint orbits in
${\cal G}^*$.
We recall that if we denote by $\mu_{\Theta}\in\tilde{\cal G}^*$ the element
we get by evaluating $\Theta$ at the origin of the group $\tilde{G}$, we
obtain a symplectic orbit in $\tilde{\cal G}^*$ passing through $\mu_{\Theta}$.
This orbit is diffeomorphic to $\tilde{G}/Ker\,d\Theta$. Therefore, we
replace the study of symplectic orbits in $\tilde{\cal G}^*$ with the
study of some quotient spaces in \Gt.
At this point,
however, instead of looking for canonical co-ordinates for
$\omega_{\mu_{\Theta}}$ or
$\Theta$, which in general do not exist globally, we use
left-invariant vector fields in \Gt, which are in the polarization,
to select an irreducible subspace of functions. These vector however, do 
not project, in general, onto the quotient space, thus they are not canonically
available in the analysis in terms of symplectic coadjoint orbit through
$\mu_{\Theta}$.


Another important advantage of the present approach is that on $\tilde{\cal
G}^*$ the enveloping algebra of $\tilde{\cal G}$ is traded with polynomial
functions and, therefore, we can only deal with their associated vector
fields via the Poisson bracket on $\tilde{\cal G}^*$, i.e. they will be
first order. This is due to the fact that Witt correspondence (see e.g.
\cite{Witt}) from the universal enveloping algebra to polynomials on
$\tilde{\cal G}^*$ is only a vector-space map, as it destroys the algebra
character.

More interesting and subtle is perhaps the comparison with the
Borel-Weyl-Bott group representation technique \cite{Pressley} intended for
finite-dimensional semisimple groups. There, the starting point is a
principal fibration of a semisimple group $G$ on the quotient $G/H$ of $G$
by the Cartan subgroup $H$, and then a condition analogous to the
polarization condition is imposed by means of the generators in a Borel
subalgebra constituted by the generators of $H$ as well as those associated
with a maximal set of positive roots. The notion of  Borel subalgebra
coincides with our definition of polarization for the case of
finite-dimensional semisimple groups, for which polarizations are always
full and symplectic (that is, finite-dimensional semisimple groups are not
anomalous). Apart from the obvious similarity, there are non-trivial
differences.  The BWB mechanism does not apply to the infinite-dimensional
case, where the Whitehead Lemma no longer holds and non-trivial cohomology
appears characterizing projective representations, as is the case of
Kac-Moody and Virasoro groups. Furthermore, these groups can be anomalous,
so that  Borel-like subalgebras do not exist. The more representative
example is constituted by the Virasoro group (see Sec. 4).

One should also add that working on $\tilde{G}$ allows us to use the left and
right universal enveloping algebras. But now we can use ideas from quantum
groups to consider higher-order polarizations in deformations of these
enveloping algebras. The main feature of their null mutual commutator is still
preserved, so that we  could have unitary representations of the deformed
right-invariant enveloping algebras. This procedure allows us to tackle the
problem of quantization of  Lie-Poisson groups without going through the
``star product quantization". We shall take up these aspects somewhere else.

As far as the geometry of anomalous systems is concerned, we want to remark
that, as commented before, the use of higher-order polarizations does not
lead, in general, to the
notion of Lagrange submanifold associated with the representation. Higher-order
polarizations are subalgebras of the left enveloping algebra of $\tilde{G}$
which are not necessarily the enveloping algebra of a given subalgebra
of $\tilde{{\cal G}}$, so that the set of solutions of the polarization
equations and equivariance condition, is a subspace
${\cal H}\subset{\cal F}^C(\tilde{G})$
which is not necessarily a subalgebra. Then, the Gelfan'd-Kolmogoroff theorem
\cite{Gelfan'd} cannot be applied to identify a submanifold of $\tilde{G}$
with the set
$Hom_C({\cal H},\; C)$. From the point of view of
Quantum Mechanics (resp. group representation) the lack
of classical integrability of polarizations makes unclear the idea of classical
limit and the association of specific phase spaces (resp. co-adjoint orbits)
with actual quantizations (resp. irreducible unitary representations). This
fact was first stated when studying the irreducible representations of the
Virasoro group ``associated" with the non-K\"{a}hler orbits
$diffS^1/SL^{(r)}(2,R),\, r>1$ \cite{Witten,Virasoro}.
More specifically, when the characteristic subalgebra in a higher-order
polarization is itself of higher order, $\GC^{HO}$, i.e. there are elements
in ${\cal P}^{HO}$ which never reproduce $X_0$, nor any power of it, under
commutation with the whole enveloping algebra, the exponential of $\GC^{HO}$
is not a subgroup of \Gt and therefore the quotient $G/{\cal G}_C^{HO}$ is
not defined in general, unlike the non-anomalous case where
$G/{\cal G}_C$, the quotient of $G$ by the integrable distribution ${\cal
G}$, constitutes the classical phase space. Furthermore, the generalized
equations of motion of higher-order type select wave functions on which the
group \Gt acts through a representation in terms of higher-order
differential operators. If the group is not anomalous, this representation
will be unitarily equivalent to a representation obtained by means of a
first-order polarization, as it happens for the Galilei group or the
harmonic oscillator group, here considered, where there exist a unitary
operator relating the representations in configuration space in terms of
higher-order differential operators with the ones in momentum space or
Bargmann-Fock space, respectively, in terms of first-order differential
operators. But if the group is anomalous, there can be representations obtained by
means of higher-order polarizations that are not unitarily equivalent to any one
in terms of first-order differential operators.

\section*{Acknowledgements}
V.A. and J.G. wish to thank the INFN, Sezione di Napoli, for financial support,
and J. G. thanks the Dipartimento di Fisica, Universit\'{a} di Napoli, for its
hospitality and the Spanish MEC for a postdoctoral grant. We all are
grateful to Januzs Grabowski for valuable discussion and for reading the
manuscript.

\pagebreak


\begin{thebibliography}{99}

\bibitem{Hove}  van Hove, J.: Acsd. R. Belg., Cl. Sci., Mem., Collect.,
                {\bf 6}, t. XXVI (1951)

\bibitem{Stone}   Stone, M.: Proc. Nat. Ac. (1929,1930)

\bibitem{Souriau}  Souriau, J.M.: {\it Structure des Sys\`{e}mes Dynamiques},
                   Dunod (1970)

\bibitem{Woodhouse} Woodhouse, N.: {\it Geometric Quantization}, Clarendon,
                    Oxford (1980)

\bibitem{23}   Aldaya, V. and de Azc\'{a}rraga, J.: J. Math. Phys. {\bf 23},
                    1297 (1982)

\bibitem{Anomalias}  Aldaya, V.,  Navarro-Salas, J.,  Bisquert, J.  and  
            Loll, R.:  J. Math. Phys. {\bf 33}, 3087 (1992)

\bibitem{Barut} Barut, A.O. and Raczka, R.: {\it Theory of Group 
         Representations and Applications}, World Scientific Publishing,
         Singapore (1986)

\bibitem{Unitariedad} Aldaya, V. and Guerrero, J., in preparation.

\bibitem{Ramirez}   Aldaya, V., Navarro-Salas, J. and Ram\'{\i} rez, A.: 
                  Commun. Math. Phys. {\bf 121}, 541 (1989)

\bibitem{Formal}  Aldaya, V. and  Navarro-Salas, J.:
                 Commun. Math. Phys. {\bf 113}, 375 (1987)

\bibitem{Saletan}  Saletan, E.J.: J. Math. Phys. {\bf 2}, 1 (1961)

\bibitem{Pseudo}  Aldaya, V. and  de Azc\'{a}rraga, J.A.: Int. J. Theo. Phys.
                 {\bf 24}, 141 (1985)

\bibitem{Bargmann}  Bargmann, V.: Ann. Math. {\bf 59}, 1 (1954)

\bibitem{Jacobson} Jacobson, N.: {\it Lie Algebras}, Dover Publications, 
                   Inc. New York (1979)

\bibitem{Position} Aldaya, V., Bisquert, J.,  Guerrero, J. and
                   Navarro-Salas, J. : J. Phys. {\bf A26}, 5375 (1993)

\bibitem{Mickelsson}  Mickelsson, J.: {\it Current Algebras and Groups},
                      Plenum Monographs in Nonlinear Physics, Plenum (1989)

\bibitem{Pressley}  Pressley A., and  Segal, G.: {\it Loop Groups}, 
                  Clarendon Press, Oxford (1986)

\bibitem{Pontrjaguin} Pontrjaguin, L.S.: {\it Topological Groups}, 
                  Princeton (1939)

\bibitem{Kirillov} Kirillov, A.A.: {\it Elements of the Theory of
                   Representations}, Springer-Verlag  (1976)

\bibitem{Virasoro}   Aldaya, V. and  Navarro-Salas, J.: Commun. Math. Phys.
                     {\bf 139}, 433 (1991)

\bibitem{Mickelsson2} Mickelsson, J.: Phys. Rev. Lett. 55, 2099 (1985)

\bibitem{Bal}  Balachandran, A.P., Marmo, G.,Skgerstam, B.S. and Stern, A.,
             {\it Classical Topology and Quantum States}, World Scientific,
              Singapore, (1991)

\bibitem{Wolf} Aldaya, V., de Azcarraga, J.A. and Wolf, K.B.,
               J. Math. Phys. {\bf 25}, 506 (1984)

\bibitem{Lang}   Lang, S.: {\it $SL_2(R)$}, Addison-Wesley P.C. (1975)

\bibitem{ConfDubna} Aldaya, V., Guerrero, J. and Marmo, G.: Int. J. Mod.
       Phys. {\bf A 12}, 3-11 (1997)

\bibitem{Bregenz}   Aldaya, V., Guerrero, J. and Marmo, G.: {\it Quantization
                    on Lie Groups},
                   Proceedings of the International Symposium Symmetries
                   and Sience X 1997, Plenum Press Corporation, New York 
                   and London, pp. 1-36 (1998)

\bibitem{Levi-Civita} Levi-Civita, T. and Amaldi, U.: {\it Lezioni di
       Meccanica Razionale}, Zanichelli, Bologna (1974) (reprinted version of
       1949 edition)

\bibitem{Frachall}   Aldaya, V.,  Calixto, M. and  Guerrero, J.:
          Commun. Math. Phys. {\bf 178}, 399 (1996)

\bibitem{Gotay}  Fern\'{a}ndez, M.,  Gotay, M.J. and  Gray, A.: Proc. Am. 
                 Math. Soc. {\bf 103}, 1209 (1988)

\bibitem{Jackiw} Jackiw, R. In: {\it Current Algebra and Anomalies},
                         Treiman, S.B.,   Jackiw, R.,  Zumino, B. and  Witten,
                         E. (eds.) World Scientific (1985)

\bibitem{Niederer}   Niederer, U.: Helv. Phys. Acta {\bf 45}, 802 (1972);
                    {\bf 46}, 191 (1973); {\bf 47}, 167 (1974)

\bibitem{Streater} Streater, R.F.: Comm. Math. Phys. {\bf 4}, 217 (1967)

\bibitem{Relativistic}
        Aldaya, V. and  Guerrero, J.: J. Phys. {\bf A28}, L137 (1995);\,
        Aldaya, V.,  Bisquert, J.,  Guerrero, J. and
        Navarro-Salas, J.: Rep. Math. Phys. {\bf 37}, 387 (1996)

\bibitem{symplin} Aldaya V., Calixto, M. and Guerrero, J.: Int. J. Mod.
        Phys. {\bf A13}, 4889 (1998)

\bibitem{Optica}  Yuen, H.P.: Phys. Rev. {\bf A 13}, 2226 (1976)

\bibitem{Kac}  Kac, V.G.: Lecture Notes in Physics {\bf 94}, 441. Berlin,
               Heidelberg, New York: Springer (1982)


\bibitem{Witt}  Humphreys, J.E.: {\it Introduction to Lie Algebras and 
                Representation Theory}, Graduate Text in Mathematics, 
                Springer-Verlag, New York (1972)

\bibitem{Gelfan'd}   Gelfand, I.M.,  Raikov, D.A. and  Chilov, G.E.: { Les
                     Anneaux normis commutatifs}, Gauthier-Villars, Paris 1964

\bibitem{Witten}    Witten, E.: Commun. Math. Phys. {\bf 114}, 1 (1988)






\end{thebibliography}
\end{document}